\documentclass[sigconf, nonacm]{acmart} 
\renewcommand\footnotetextcopyrightpermission[1]{} 
\usepackage[nolist]{acronym}
\usepackage{listings}
\usepackage{array}
\usepackage{bm}
\usepackage{enumitem}
\usepackage[toc,page]{appendix}
\usepackage{amsmath}
\usepackage{float}
\usepackage{algpseudocode}
\usepackage[ruled]{algorithm2e}
\usepackage{tikz,tikz-qtree,tikz-qtree-compat}
\usepackage{caption}
\usepackage{subcaption}
\usepackage{pgfplots}
\usepackage{tabularx}
\usepackage{multirow}
\usepackage{booktabs}
\usepackage{setspace}
\usepackage{makecell}
\usepackage{todonotes}
\usepackage{pdfpages}

\usepackage{dsbda-style}

\newsavebox{\twosubbox}

\pgfplotsset{compat=newest}

\newcommand{\thesistitle}{Reducing a Set of Regular Expressions and Analyzing Differences of Domain-specific Statistic Reporting}

\newcolumntype{L}{>{\raggedright\arraybackslash}X}
\newcolumntype{V}[1]{>{\raggedright\arraybackslash}m{#1}}
\newcolumntype{C}{>{\centering\arraybackslash}X}
\newcolumntype{M}[1]{>{\centering\arraybackslash}m{#1}}

\newcommand\YAMLcolonstyle{\ttfamily\color{red}\mdseries}
\newcommand\YAMLkeystyle{\ttfamily\color{black}\bfseries\small}
\newcommand\YAMLvaluestyle{\ttfamily\color{blue}\mdseries}

\makeatletter
\def\NAT@spacechar{~}
\makeatother

\makeatletter

\newcommand\language@yaml{yaml}

\expandafter\expandafter\expandafter\lstdefinelanguage
\expandafter{\language@yaml}
{
  keywords={true,false,null,y,n},
  keywordstyle=\color{darkgray}\bfseries,
  basicstyle=\YAMLkeystyle,                                 
  sensitive=false,
  comment=[l]{\#},
  morecomment=[s]{/*}{*/},
  commentstyle=\color{purple}\ttfamily,
  stringstyle=\YAMLvaluestyle\ttfamily,
  moredelim=[l][\color{orange}]{\&},
  moredelim=[l][\color{magenta}]{*},
  moredelim=**[il][\YAMLcolonstyle{:}\YAMLvaluestyle]{:},   
  morestring=[b]',
  morestring=[b]",
  literate =    {---}{{\ProcessThreeDashes}}3
                {>}{{\textcolor{red}\textgreater}}1     
                {|}{{\textcolor{red}\textbar}}1 
                {\ -\ }{{\mdseries\ -\ }}3,
}

\lst@AddToHook{EveryLine}{\ifx\lst@language\language@yaml\YAMLkeystyle\fi}
\makeatother

\newcommand\ProcessThreeDashes{\llap{\color{cyan}\mdseries-{-}-}}

\newcommand{\regex}[1]{\textbf{\texttt{#1}}}

\begin{document}
\pagestyle{plain}
\title{\thesistitle}

\author{Tobias Kalmbach, Marcel Hoffmann, Nicolas Lell, Ansgar Scherp}
\email{{firstname.lastname}@uni-ulm.de}
\affiliation{
  \institution{Data Science and Big Data Analytics Group, Ulm University}
  \country{Germany}
}

\begin{acronym}
    \acro{APA}[APA]{American Psychology Association}
    \acro{CORD}[CORD-19]{COVID-19 Open Research Dataset}
    \acro{DFA}[DFA]{Deterministic Finite Automaton}
    \acro{HCI}[HCI]{Human-Computer-Interaction}
    \acro{NFA}[NFA]{Nondeterministic Finite Automaton}
    \acro{SYA}[SYA]{shunting-yard algorithm}
\end{acronym}

\newlist{questions}{enumerate}{2}
\setlist[questions,1]{label=\textbf{RQ\arabic*.},ref=RQ\arabic*}
\setlist[questions,2]{label=(\alph*),ref=\thequestionsi(\alph*)}
\setlist[questions]{topsep=0pt}

\newcommand{\kanonymity}{\mbox{$\mathscr{k}$-anonymity}}
\newcommand{\kmanonymity}{\mbox{$\mathscr{k}^m$-anonymity}}
\newcommand{\ldiversity}{\mbox{$\mathscr{l}$-diversity}}
\newcommand{\tcloseness}{\mbox{$\mathscr{t}$-closeness}}
\newcommand{\cldiversity}{\mbox{(c,$\mathscr{l}$)-diversity}}
\newcommand{\edifprivacy}{\mbox{$\mathscr{\varepsilon}$-differential privacy}}



\begin{abstract}
Due to the large amount of daily scientific publications, it is impossible to manually review each one.
Therefore, an automatic extraction of key information is desirable. 
In this paper, we examine STEREO, a tool for extracting statistics from scientific papers using regular expressions.
By adapting an existing regular expression inclusion algorithm for our use case, we decrease the number of regular expressions used in STEREO by about $33.8\%$.
We reveal common patterns from the condensed rule set that can be used for the creation of new rules.
We also apply STEREO, which was previously trained in the life-sciences and medical domain, to a new scientific domain, namely \acf*{HCI}, and re-evaluate it.
According to our research, statistics in the HCI domain are similar to those in the medical domain, although a higher percentage of APA-conform statistics were found in the HCI domain.
Additionally, we compare extraction on PDF and \LaTeX{} source files, finding \LaTeX{} to be more reliable for extraction.
\end{abstract} 

\settopmatter{printacmref=false,printfolios=true}
\maketitle
\pagestyle{plain}

\section{Introduction}
\label{sec:introduction}
An abundance of scientific papers are published daily. 
The large and rapidly growing number of papers is too extensive to scan manually.
In particular, assessing the published results in terms of insights generated by the statistical analyses is very challenging. 
A quick overview of published statistics can be used by authors to find statistical errors in their studies more easily or improve their notation.
Additionally, readers can easily access statistics to verify and check them.
Moreover, extracting sentences containing statistics together with metadata (authors, title, etc.) can enable researchers to get an impression of an article without the need to fully read it.
Reliably knowing if and optimally what statistics a paper publishes can also be important for further research or usage in new papers, such as related work.
Tools like \emph{statcheck}~\cite{nuijten2016} provide very accurate extraction of statistics reported in accordance with the commonly used writing style guide of the \ac{APA}~\cite{bentley1929}.
However, previous research found that in a sample of $113$k statistics, extracted from pre-prints in the life sciences and medical domain, less than one percent were \ac{APA}-conform~\cite{epp2021}.
In this work, we extend STEREO (STat ExtRaction Experimental cOnditions)~\cite{epp2021}, an automatic statistics extraction pipeline for statistics presented in \ac{APA} style notation as well as non-\acs{APA} notation.
STEREO learns regular expressions (rules) to decide whether a sentence contains statistics ($R^+$ rules) or not ($R^-$ rules) using active wrapper learning.
The $R^+$ rules are used to extract the statistic's type and values.
During the development of STEREO, $85$ $R^+$ rules and $1,425$ $R^-$ rules were learned.
Inspecting these rules shows that rules which were added later in the learning process generalize better and previously added rules become obsolete.
It is of interest how many of these rules are covered by others. 
Reducing the number of rules can help to identify common patterns of non-\ac{APA} statistics, e.\,g., incomplete reporting, and derive recommendations to improve statistics reporting.
Subsequently, general rule patterns that indicate a sentence does not contain a statistic can serve as guidance for future active wrapper approaches.
One can make use of these general rule patterns to avoid creating excess rules.
With this reasoning, we apply a DFA-based (\acl{DFA}) algorithm~\cite{ChenX20} to minimize the existing set of regular expressions.

The current implementation of STEREO was trained on the \ac{CORD}, a dataset containing papers on COVID-19 and coronaviruses in general.
Inspecting other scientific domains may result in finding uncovered statistic types and reporting styles as well as different variations of non-APA reporting or deviations from existing errors.
Therefore, we aim to transfer STEREO to a new domain of science, namely \acl{HCI}.
Thus, one needs to extend the rule set with additional rules. 
This growing rule set motivates a size reduction of the rule set, as well.

STEREO currently uses the JSON files from \ac{CORD}, which contain the extracted text parts of scientific papers.
In this work, we extend STEREO to support new statistic types.
We also extend STEREO to support statistic extraction from LaTeX files.
We investigate the impact the different input formats have for statistics extraction.

In summary, this work makes the following contributions:
\begin{itemize}
  \item We analyze the extraction rules from STEREO, aim to reduce the large rule set, and analyze general patterns that identify sentences that do not contain statistics. 
  These patterns can be used in future statistic extraction to effectively identify false positive statistics.
  \item We extend the rule set by repeating the active wrapper learning on a new domain, \acl{HCI}. 
  We evaluate the reduced rule set using precision.
  We also measure the runtime needed to apply the reduced rule set compared to the full rule set.
  \item We investigate the extraction from LaTeX vs. PDF files converted to text to measure statistic extraction performance and false positive counts.
\end{itemize}

Section~\ref{sec:related_work} presents related work. Section~\ref{sec:method} specifies the computation of the rule set minimization, and Section~\ref{sec:experiments} explains the experimental apparatus.
Sections~\ref{sec:results} and~\ref{sec:discussion} present and discuss the results.

\section{Related Work}
\label{sec:related_work}
First, we present some regular expression inclusion algorithms. 
Then, we give an overview of approaches for the extraction of statistics, and finally, we briefly summarize.

\subsection{Rule Set Inclusion Algorithms}
\label{sec:rw_inclAlgs}
We give a concise overview of the formal definition of regular expressions we use in this paper, following the definition of \citet{ChenX20}.
\begin{itemize}
    \item $a,b,c,...$: members of the alphabet used in the expression
    \item $\epsilon$: an empty string
    \item $ab$ or $a\&b$: the concatenation of $a$ and $b$ (i.\,e., $a$ must be directly followed by $b$)
    \item $a|b$: the choice of $a$ or $b$ (i.\,e., either $a$ or $b$ is accepted). Other literature often uses $+$ as choice operator
    \item $a^*$: the Kleene Star operator (i.\,e., $a$ can be repeated $0$ or more times)
\end{itemize}
Regarding regular expression inclusion, i.\,e., minimal rule set computation, many algorithms are limited to determining inclusion using one-unambiguous regular expressions.
One-unambiguous regular expressions were first presented by \citet{bruggemann1998one} and are a real subset of regular expressions in general. 
They are regular expressions that can match every word (in their respective language) in a unique way, without looking ahead.
For example, $(a_1|b_1)^*(a_2 | \epsilon)$ (numbered for clarity) is not one-unambiguous, as the word $baa$ can be formed as $b_1a_1a_1$ or $b_1a_1a_2$. 
However, $(b^*a)^*$ describes the same language but is one-unambiguous. 

\citet{ChenX20} presented two algorithms for regular expression inclusion.
The first is an automata-based algorithm that converts the given one-unambiguous regular expressions into \acp{DFA} and subsequently compares those.
A \ac{DFA} is a finite automaton that accepts or rejects an input by traversing states in the automaton. An input is accepted if the process terminates in an accepting state and rejected otherwise.
The second algorithm is a derivative-based algorithm. 
Derivatives of regular expressions are sub-expressions, which in turn are valid regular expressions themselves.
The idea is that if an expression A is included in an expression B, all derivatives of A are also derivatives of B.
The algorithm generates all derivatives of both expressions.
If all derivatives of one expression are included in the other, the first expression is included in the second.

Similar to the first algorithm of \citet{ChenX20}, \citet{nipkow2014} presented a framework to determine if two given regular expressions are equivalent. 
Equivalence is a stricter requirement than in the paper by \citet{ChenX20}. 
The presented framework dynamically creates an automata from one regular expression and using ``computations on regular expression-like objects''~\cite[p. 2]{nipkow2014} as a substitute for the traditional transition table. ``Regular expression-like'' means the embedding of the regular expression in a state object represented by the language of the regular expression.
Chen and Xu's algorithm has a quadratic runtime complexity with respect to the length of the regular expressions.

A different approach is presented by \citet{Hovland12}. 
Instead of using automata, an inference system is used to inductively determine a binary relationship between one-unambiguous regular expressions.
This inference system abstracts types of regular expressions and simplifies them for further use. 
If all simplifications reach the trivial case $\epsilon \subseteq l$, where $l$ is any regular expression, the two regular expressions are in an inclusion relationship.
The algorithm uses a depth-first search in combination with the inference system. 
The algorithmy by Hovland is guaranteed to run in polynomial time.

\subsection{Statistics Extraction}
As mentioned in the introduction, we aim to extract reported statistics from scientific papers.
\citet{teja2021} present a regular expression-based approach to extract statistics from scientific papers. 
They use a single regular expression to match the $p$-value in its representation according to the statistical test, e.\,g., one regular expression to match a $t$-test reported as \regex{t(df)=float, p (<, >, =) float}. 
Their model achieved an extraction accuracy of $90.2\%$ when no accompanying test statistic was present (only a $p$-value) and $79.0\%$ with a test statistic.

A similar, generally well-known approach is \emph{statcheck}~\cite{nuijten2016}.
\emph{statcheck} is an R package that allows the user to extract and verify published statistics.
\emph{statcheck} is limited to extracting statistics that match \ac{APA} guidelines exactly. 
However, this limitation always allows recalculation as well as an assessment of the consistency of the $p$-value, as all reported statistics and degrees of freedom must be present by construction.
In 2017, \citet{schmidt_2017} disputed the effectiveness of \emph{statcheck} in their paper aptly named ``Statcheck does not work: All the numbers. Reply to Nuijten et al. (2017)''. 
Here they criticize the \emph{statcheck} authors for the testing conducted in their follow-up paper~\cite{nuijten2017}. 
Upon retesting, \citet{schmidt_2017} finds that \emph{statcheck} simply does not detect many reported statistics and therefore does not flag the statistical tests. 
All in all, the reported accuracy and overall performance of \emph{statcheck} are called into question, even though \emph{statcheck} is still widely used~\cite{freedman2017, sakaluk2018, psychopen2017}.

Both \emph{statcheck}~\cite{nuijten2016} and \citet{teja2021} do not have any built-in feature to match non-\ac{APA} statistics and can therefore only match the very strict pattern.
In order to expand this strict pattern, \citet{boschen2021} uses a similar, yet more sophisticated approach than \emph{statcheck}.
They present a text-mining approach on XML documents that are structured following the Journal Archiving Tag System NISO-JATS. 
The algorithm differentiates between three result types.
Computable results, where the $p$-value is given and can be recalculated; checkable results, where the $p$-value is not given but can be calculated; and lastly, uncomputable results, where the $p$-value may or may not be reported, but cannot be calculated due to missing information.
The algorithm uses CERMINE~\cite{tkaczyk2015} to format papers. 
CERMINE (Content ExtRactor and MINEr) extracts the contents and metadata from papers given as PDFs and formats them in accordance with NISO-JATS.
With this structure, \citet{boschen2021} applies several transformations to letters and numbers, e.\,g., greek letters or fractions.
The authors note that with just these transformations, the original \emph{statcheck} performs better, due to fewer missed statistical tests.
After these transformations, the extraction algorithm by \citet{boschen2021} works as follows: Sentences are only selected if they contain at least one letter, followed by an operator ($<,>,=,\leq,\geq$), which in turn is followed by a number.
This way, the sentences are split at a set of given words (e.\,g., and, or, of,...) and at commas following a word. 
Surrounding text is removed using regular expressions. 
Individual heuristics, to cope with varying reporting styles, are applied to extract the recognized test statistics, the operator, degrees of freedom, and $p$-value.
As the requirements are not as strict as \emph{statcheck}'s, \citet{boschen2021} generally achieves a higher accuracy.

Lastly, STEREO~\cite{epp2021} uses active wrapper learning to learn regular expressions (rules) that determine whether or not a sentence contains statistics. 
The rules are divided into $R^+$ and $R^-$ rules. 
$R^+$ rules are rules that match statistics presented in a sentence, while $R^-$ rules denote that a sentence does not include statistics. 
$R^+$ rules have additional sub-rules, which are used to capture statistic parts (e.\,g., $p$-value) after the statistic type has been identified by the main $R^+$ rule.
During the active wrapper learning, every sentence that does not contain a number is ignored.
For any remaining sentences not matched by any rules, the user is prompted to create a new rule to cover the new case.
STEREO achieved a precision close to $100\%$ for \acs{APA}-conform statistics and a precision of $95\%$ for non-\acs{APA} statistics.

\subsection{Summary}
All in all, there are already many approaches for both regular expression inclusion and statistics extraction.
As previously mentioned, this work is based on STEREO and therefore uses methods for statistics extraction presented by \citet{epp2021}. 
In particular, we transfer STEREO to a new scientific domain and potentially find new rules for statistics extraction.
Additionally, we use the automaton-based algorithm presented by \citet{ChenX20} to find a minimal rule set, as it is the most straightforward implementation and the runtime is negligible, as our regular expressions are usually short ($<100$ characters).

\section{Computing the Rule Set Inclusion}
\label{sec:method}
In Section \ref{sec:rminusrules}, we explain the benefits of $R^-$ rules.
Section \ref{sec:algdesc} details our inclusion algorithm, which tests two regular expressions for inclusion using the algorithm given by \citet{ChenX20}. 
In a preprocessing step, we convert Python regular expressions into expressions following the formal definition presented in Section~\ref{sec:rw_inclAlgs}. 
This is not part of the algorithm of \citet{ChenX20}, since they formulate their algorithm for regular expressions already following the formal definition.
As far as we are aware, we are the first to implement such a transformation.
Section \ref{sec:rule_inclusion_example} shows the algorithm with a simple example.

\subsection{Why \texorpdfstring{$R^-$}{R-} rules?}\label{sec:rminusrules}
STEREO applies many $R^-$ rules ($1,425$), while not using many $R^+$  rules ($85$).
The reasons supporting the use of these $R^-$ rules is discussed here.

Extracting text parts that follow a predefined pattern or system, in this context \acs{APA}-conform statistics, is very simple. 
The pattern produces finitely many base-structures, which can then in turn be matched.
This method was used by \citet{nuijten2016}.

Due to lack of knowledge or proficiency, not all statistics in papers are reported in \ac{APA} fashion, but with deviations ranging from slight spelling errors to disregarding the format completely.
In contrast, these deviations are in principal arbitrary, and finding rules that exactly capture these deviations is very difficult or impossible in a large-scale dataset.
As a solution to this problem, \citet{epp2021} use an adapted version of statistic recognition and extraction.
The rules that match statistics ($R^+$) are extended to be more lenient, thus accepting close matches or common mistakes. 
As more structures need to be matched, STEREO has more matching rules ($85$ for statistics detection with $52$ sub-rules for value extraction) in comparison to the matching rules in \emph{statcheck} ($8$ for detection, $23$ for value extraction).

However, during the active wrapper learning phase, when searching for statistics not yet captured by $R^+$ rules, all sentences containing numbers are marked as potential statistics, which is a very wide cast net.
In order to eliminate the need to manually investigate every non-statistic sentence, STEREO adds $R^-$ rules which determine that a sentence does not report statistics.
For example, these rules capture a number as a part of a table reference, i.\,e., ``Table 1'', and eliminate it from further processing.

A reduction of these rules can reveal common patterns that can provide a good guideline on how to write $R^-$ rules for similar future active wrapper approaches.
Moreover, a size reduction will result in a faster runtime as fewer rules need to be applied for each sentence.

\subsection{Algorithm Description}\label{sec:algdesc}
In the following, we describe the algorithm for the inclusion of regular expressions.
Each given regular expression is preprocessed.
Then a \ac{DFA} is constructed for comparison, following the inclusion algorithm given by \citet{ChenX20}.
At first, it might seem easy to apply an inclusion algorithm to regular expressions directly. 
This is not necessarily the case. 
Regular expressions, especially Python regular expressions, can represent the same thing in many ways.
For example, a three letter word containing \regex{a}, \regex{b}, \regex{c} in any order and quantity can be represented as \regex{[abc]\{3\}}, \regex{(a|b|c)\{3\}}, \regex{[abc][abc][abc]}, and more.
Therefore, applying an inclusion algorithm to regular expressions directly requires more differentiation.
As mentioned in Section \ref{sec:related_work}, \citet{Hovland12} uses an inference system to test regular expression inclusion directly.
In their algorithm, they use ten different cases to abstract a given regular expression until the inclusion problem is easily decidable.

In general, every regular expression can be converted into an equivalent \ac{DFA}.
Using \acp{DFA}, it is possible to apply the same algorithm state-by-state without the need to differentiate between different cases.
The algorithm presented by \citet{Hovland12} is more time efficient than an algorithm~\cite{ChenX20} using the construction of equivalent automatons.
However, the implementation of the latter is simpler and the efficiency is not of utmost importance in our case. 
The regular expressions we work with are usually less than $100$ characters long before preprocessing, with a few exceptions\footnote{$16$ regular expressions were longer than $100$ characters and $3$ regular expressions were longer than $150$ characters (longest $225$ characters)}, making the performance difference mostly negligible.
Therefore, we use the algorithm described by \citet{ChenX20}. 
The notation we use for automata is defined by~\cite{LucasR05}:
\begin{itemize}
    \item $A$: an automaton
    \item $Q_A$: set of all states in $A$
    \item $q_A$: start state in $A$
    \item $F_A$: set of all accepting states in $A$
    \item $\delta_A$: all transitions in $A$
    \item $\Sigma_A$: alphabet used in $A$
    \item $L(A)$: the language accepted by $A$
\end{itemize}

\begin{algorithm}
\LinesNumbered
\caption{Regular Expression Inclusion}\label{alg:inclusion}
\KwData{List of regular expressions $R$}
\KwResult{Array detailing every regular expression included by an other}
Initialize array $S$ with length of regular expression list\;
\For{$r_1$ in $R$}{
    Initialize empty set $T$\;
    $c_1 \gets \text{normalize}(r_1)$\;
    $N_1 \gets \text{nfa}(c_1)$\;
    $A_1 \gets \text{dfa}(N_1).\text{asComplete}()$\;
    
    \For{$r_2$ in $R, r_2 \neq r_1$}{
        $c_2 \gets \text{normalize}(r_2)$\;
        $N_2 \gets \text{nfa}(c_2)$\;
        \If{$\Sigma_{N_2} \nsubseteq \Sigma_{N_1}$}{
            continue with next $r_1$\;
        }
        $A_2 \gets \text{dfa}(N_2).\text{asComplete}()$\;
        \tcc{Inclusion check with Algorithm \ref{alg:optimized_inclusion}}
        \If{$A_1.\text{inclusion}(A_2)$}{ 
            Add $r_2$ to $T$\;
        }
    }
    Save $T$ at the index of $r_1$ in $S$\;
}
\end{algorithm}

Our adapted implementation can be seen in Algorithm~\ref{alg:inclusion}. 
The algorithm can be split into two steps, which are repeated for every rule.
In the first step, a regular expression is preprocessed and an equivalent, complete DFA is created. 
This regular expression is the rule for which we want to find all expressions it includes. 
Next, excluding the rule used in step one, we iterate through each rule in the set of rules.
We use the same preprocessing and automaton creation for these rules.
The resulting automata are subsequently checked for inclusion.
The algorithm's output is an array of lists, where each position in the array corresponds to a rule's index in the given list of regular expressions.
The respective lists in the array contain all the rules included by the rule at that index.

\paragraph{Preprocessing}
We begin by transforming the given Python regular expression to a formal definition by preprocessing them.
For readers not familiar with the syntax of Python regular expressions, we recommend the official documentation\footnote{\url{https://docs.python.org/3/howto/regex.html}} as a reference.
In the first step, we replace all meta characters (e.\,g., \regex{\textbackslash d} or \regex{\textbackslash w}) with their expanded form (e.\,g., \regex{( 0 | 1 | 2 | ... | 8 | 9 )}). 
Also, character sets (e.\,g., \regex{[a-z]}) and repetition qualifiers are expanded as well, e.\,g., \regex{a\{2,4\}} becomes \regex{aa|aaa|aaaa}. 
Additionally, Python does not have a symbol for concatenation, so we fill all implicit concatenation with an \regex{\&} for the next step, e.\,g., \regex{ab} becomes \regex{a\&b}.
Next, we use Dijkstra's \ac{SYA}~\cite{dijkstra1961} for a definitive normalized form that does not include any parentheses.
The \ac{SYA} transforms expressions, in this case, regular expressions, from infix notation to postfix notation, e.\,g., \regex{a|b} becomes \regex{ab|}.
In short, \ac{SYA} uses an output and operator stack and a predefined operator precedence. 
For regular expressions, the operator precedence is '$*$', '$\&$', '$|$' from highest to lowest.
The given expression is read character by character.
All elements of the alphabet are immediately pushed onto the output stack, while operators are pushed to the operator stack. 
If the operator stack is non-empty, all contained operators are popped off the operator stack and added to the output stack while the top-most operator has an equal or higher precedence. 
Once an operator with lower precedence is on top of the operator stack, add the regarded operator to the operator stack.
A special case is parentheses. 
When finding an opening bracket, it is always added to the operator stack.
The normal procedure is continued until the closing bracket is found, upon which all operators are popped and added to the output, until the corresponding opening bracket is found.
The parentheses are then discarded.
\extended{We show the application step-by-step at an example in Table \ref{tab:shunting_yard}.
Another thorough example of an application of \ac{SYA} is given by \citet[p. 3]{rastogi2015}. 
In this example, the algorithm is applied to an arithmetic example, which also demonstrates the procedure.}

\begin{figure}[t]
     \begin{subfigure}[t]{\linewidth}
         \centering
         \includegraphics[width=\textwidth]{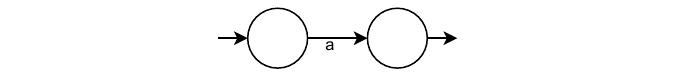}
         \caption{Simple NFA created for a symbol $a\in\Sigma \cup \epsilon$}
         \label{fig:tc_simple}
     \end{subfigure}
     \vspace{.2cm}
     \begin{subfigure}[t]{\linewidth}
         \centering
         \includegraphics[width=\textwidth]{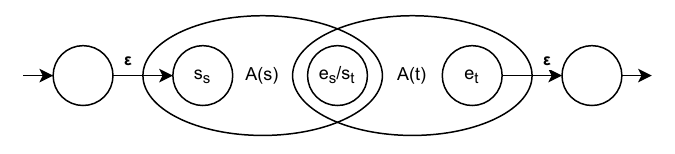}
         \caption{NFA created for the expression $st$}
         \label{fig:tc_concat}
     \end{subfigure}
     \vspace{.2cm}
     \begin{subfigure}[t]{\linewidth}
         \centering
         \includegraphics[width=\textwidth]{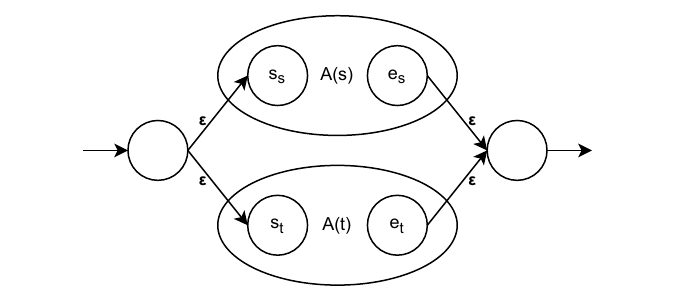}
         \caption{NFA created for the expression $s|t$}
         \label{fig:tc_alt}
     \end{subfigure}
     \vspace{.2cm}
     \begin{subfigure}[t]{\linewidth}
         \centering
         \includegraphics[width=\textwidth]{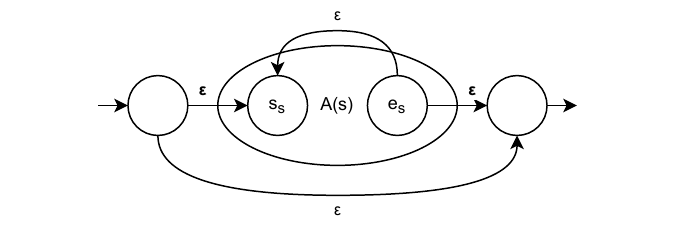}
         \caption{NFA created for the expression $s*$}
         \label{fig:tc_kleene}
     \end{subfigure}
        \caption[Thompson's construction - General]{Basic NFAs used in Thompson's Construction. $s,t$ are regular (sub-)expressions. $A(s)$ refers to the automaton created for $s$. $s_s$ and $e_s$ refer to the starting and final state of $A(s)$ respectively. $\epsilon$ is an empty transition.}
        \label{fig:tc_basic_nfas}
\end{figure}
\paragraph{Automaton Construction} 
As mentioned in Section~\ref{sec:related_work}, \citet{ChenX20} present two approaches, of which we use the automaton-based algorithm.
When given two regular expressions, they convert these into Glushkov automata, using Glushkov's construction~\cite{glushkov1961}, and check the inclusion of the resulting \ac{DFA}.
In our implementation we do not use Glushkov's construction but Thompson's construction~\cite{Thompson68}, which creates an \ac{NFA}, since Thompson's constructions can be implemented in a more straight forward way.
This \ac{NFA} has to be converted to a \ac{DFA}.
However, the resulting \ac{DFA} from the \ac{NFA} have been shown to be equivalent~\cite{sakarovitch_2009}.
Thompson's construction~\cite{Thompson68} builds the \ac{NFA} with a bottom-up approach.
Given a regular expression, first NFAs are constructed for the sub-expressions and then merged.
The NFAs for sub-expression do not necessarily have a start or accepting state and may have non-connected arrows, which will be connected when merging.
In Thompson's construction, there are four basic NFAs for a simple transition with a given letter, concatenation, e.\,g., $ab$, alternative, e.\,g., $a|b$, and the Kleene star expression, e.\,g., $a*$.
A general representation of these basic NFAs can be found in Figure~\ref{fig:tc_basic_nfas}.
This representation is adapted from \citet{AhoSU86}.
Some more complete overviews of the construction are presented by \citet{Watson93} and \citet{AhoSU86}.

After applying Thompson's construction to a preprocessed regular expression, we convert the resulting \ac{NFA} into a \ac{DFA} with powerset construction~\cite{Rabin59}.

In the following, we further discuss the inclusion algorithm used to determine $A_1.\text{includes}(A_2)$ (see Algorithm~\ref{alg:inclusion}). 

\begin{algorithm}
\SetNoFillComment
\LinesNumbered
\caption{Automaton inclusion algorithm by \citet{ChenX20}.
\extended{
This Algorithm~\ref{alg:optimized_inclusion} is an optimized version of Algorithm~\ref{alg:aut_inclusion} as shown in the Appendix~\ref{sup:automaton_inclusion}.}
}
\label{alg:optimized_inclusion}
\KwData{Two \acp{DFA}, $A'$ the complement of $A_1$ and $A_2$}
\KwResult{Boolean whether $A_2 \subseteq A_1$}
Initialize empty stack $Q$\;
Push $(q_{A'}, q_{A_2})$ onto $Q$\;
\While{$Q$ is not empty}{
  $(p, q) \gets Q.pop()$\;
  \If{$(p, q)$ is unmarked}{
        \For{$a$ in $\Sigma$}{
            \If{$\delta_{A_2}(q, a)$ is defined}{
                \eIf{$\delta_{A'}(p, a)$ and $\delta_{A_2}(q, a)$ are accepting states}
                {
                    return $FALSE$\;
                }{
                    Push $(\delta_{A'}(p, a), \delta_{A_2}(q, a))$ onto $Q$\;
                    Mark $(p, q)$\;
                }
            }
            
        }
  }
}
\Return $TRUE$\;
\end{algorithm}

\paragraph{Inclusion Algorithm}
Following \citet{ChenX20}, we require $\Sigma_{r2} \subseteq \Sigma_{r1}$ as a necessary condition for a regular expression $r1$ to include another regular expression $r2$.
This is already checked in Algorithm~\ref{alg:inclusion}.
The specific inclusion algorithm for two \acp{DFA} is shown in Algorithm~\ref{alg:optimized_inclusion}.
This is an optimized algorithm where we iterate state-by-state using the same transition letter for both automata, keeping track of our current position with a tuple $(p,q)$.
The algorithm terminates as soon as we reach a goal state in both automata, returning earlier than without the optimization.

The intuitive reasoning behind the inclusion algorithm is that the complement of an automaton accepts the ``exact opposite'' language, i.\,e., all previously accepted words are not accepted and vice versa. 
We use an example automaton called $A_1$ and its complement $A'$.
When combining $A'$ with a different automaton, $A_2$, a path from the beginning state to an accepting state can only exist when the language of the combined automaton is non-empty.
In other words, $A_2$ accepts a word that $A'$ also contains in their language.
$A_2$ accepts a word that $A_1$ does not, so $A_2$ is not included since an automaton and its complement cannot accept the same word.

The combination of the two automata $A'$ and $A_2$
is done by constructing an automaton with the cross product of the respective states (i.\,e., $Q_{A'} \times Q_{A_2}$), merging the alphabets (i.\,e., $\Sigma = \Sigma_{A'} \cup \Sigma_{A_2}$), and defining the start state $q$ as the tuple $(q_{A'}, q_{A_2})$ and the accepting states $F=F_{A'} \times F_{A_2}$. 
Furthermore, we define the two-fold transition function $\delta((p,q), a) = (\delta_{A'}(p, a), \delta_{A_2}(q, a))$.
We can always assume that $\delta_{A'}(q, a)$ exists if $\delta_{A_2}(p, a)$ does, as $\Sigma_{A_2} \subseteq \Sigma_{A_1}$ is required and checked, and $A'$ and $A_2$ are complete automata.

All in all, the algorithm checks whether $\overline{L(A_1)} \cap L(A_2) = \emptyset$, which is exactly non-empty if the language of $A_2$ contains some word that is not in the language of $A_1$.

\subsection{Algorithm Example}\label{sec:rule_inclusion_example}
\begin{figure}[t]
    \centering
    \includegraphics[width=\linewidth]{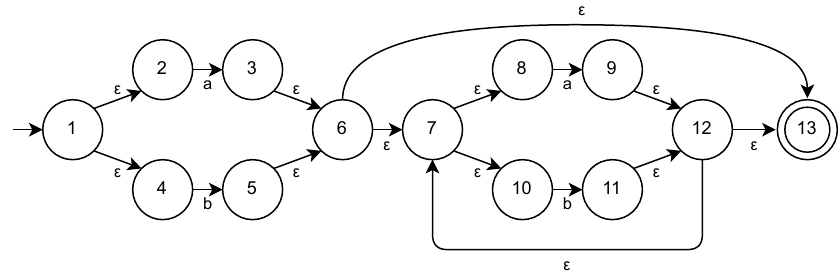}
    \caption[Thompson's construction for \texttt{[a-b](a|b)*}]{Thompson's construction for \regex{ba|ab|*\&} (input was \regex{[a-b](a|b)*}). Nodes $1$ through $6$ represent \regex{[a-b]}, while nodes $6$ through $13$ represent \regex{(a|b)*}. For a larger version see Appendix~\ref{sup:tc_example}.}
    \label{fig:tc_example}
\end{figure}

\begin{figure}[t]
    \centering
    \includegraphics[width=\linewidth, height=4cm, keepaspectratio]{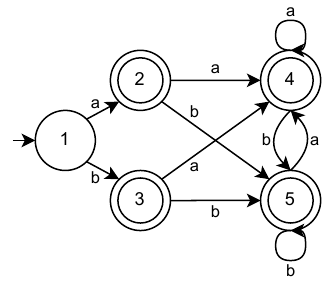}
    \caption[Powerset construction for \texttt{[a-b](a|b)*}]{Powerset construction for \regex{ba|ab|*\&} (input was \regex{[a-b](a|b)*}). Note: This is not an optimally minimized DFA.}
    \label{fig:tc_dfa}
\end{figure}

To illustrate our approach, we use the regular expression \regex{[a-b] (a|b)*} as an example.
We start with escaping our expression, resulting in \regex{(b|a)\&(a|b)*}.
Continuing with the \acl{SYA}, we get \regex{ba|ab|*\&}.
\extended{A step-by-step conversion for our example can be seen in Table~\ref{tab:shunting_yard}.}
The Thompson's construction for our example can be seen in Figure~\ref{fig:tc_example} and the subsequent powerset construction is found in Figure~\ref{fig:tc_dfa}.

\extended{
\begin{table}
\caption[Step-by-step shunting-yard conversion of \texttt{(b|a)\&(a|b)*}]{Step-by-step shunting-yard conversion of \regex{(b|a)\& (a|b)*}. The reason column indicates what rule was used to get the \emph{next} line. $\Sigma$ refers to the alphabet of the expression, 'op.' is short for operator, and $>$ refers to higher precedence.}
\label{tab:shunting_yard}
\small
\begin{tabular}{l|c|c|l|l}
\thead{Input}   & \thead{Regarded\\ Letter or\\ Operator} & \thead{Operator\\ Stack} & \thead{Output \\ Stack} & \thead{Reason} \\ \hline
$(b|a)\&(a|b)*$ & -               & -              & -              & -                 \\
$b|a)\&(a|b)*$  & $($             & -              & -              & Opening (         \\
$|a)\&(a|b)*$   & $b$             & $($            & -              & $b \in \Sigma$    \\
$a)\&(a|b)*$    & $|$             & $($            & $b$            & op.               \\
$)\&(a|b)*$     & $a$             & $(|$           & $b$            & $a \in \Sigma$    \\
$\&(a|b)*$      & $)$             & $(|$           & $ba$           & Closing )         \\
$\&(a|b)*$      & $)$             & $($            & $ba|$          & Closing )         \\
$(a|b)*$        & $\&$            & -              & $ba|$          & op.               \\
$a|b)*$         & $($             & $\&$           & $ba|$          & Opening (         \\
$|b)*$          & $a$             & $\&($          & $ba|$          & $a \in \Sigma$    \\
$b)*$           & $|$             & $\&($          & $ba|a$         & op.               \\
$)*$            & $b$             & $\&(|$         & $ba|a$         & $b \in \Sigma$    \\
$*$             & $)$             & $\&(|$         & $ba|ab$        & Closing )         \\
$*$             & $)$             & $\&($          & $ba|ab|$       & Closing )         \\
-               & $*$             & $\&$           & $ba|ab|$       & op., $* > \&$     \\
-               & -               & $\&*$          & $ba|ab|$       & Pop op.           \\
-               & -               & $\&$           & $ba|ab|*$      & Pop op.           \\ 
-               & -               & -              & $ba|ab|*\&$    & -
\end{tabular}
\end{table}
}

\begin{figure}[t]
     \begin{subfigure}[t]{\linewidth}
         \centering
         \includegraphics[width=\textwidth, height=4cm, keepaspectratio]{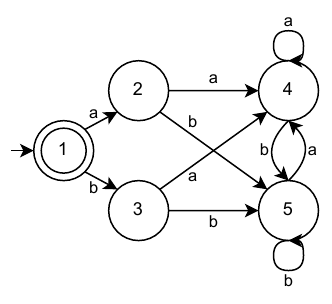}
         \caption[Complement of the DFA created for the expression \regex{[a-b](a|b)*}]{Complement of the DFA created for the expression \regex{[a-b](a|b)*} (see Figure \ref{fig:tc_dfa}).}
         \label{fig:inc_a1}
     \end{subfigure}
     \begin{subfigure}[t]{\linewidth}
         \centering
         \includegraphics[width=\textwidth, height=4cm, keepaspectratio]{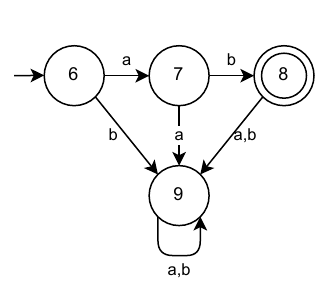}
         \caption{The complete DFA created for the expression \regex{ab}}
         \label{fig:inc_a2}
     \end{subfigure}
    \caption[Inclusion Algorithm - Example]{The DFAs created for the expressions \regex{[a-b](a|b)*} and \regex{ab} respectively. Note that the DFA for \regex{[a-b](a|b)*} has been converted to the complement and both DFAs are complete.}
    \label{fig:inclusion_automaton}
\end{figure}

We now show how the improved inclusion algorithm works by using a simple example.
We check if \regex{ab} $\subseteq$ \regex{[a-b](a|b)*}, which is true.
As the \ac{DFA} for \regex{[a-b](a|b)*} is complete already, we only need to form the complement.
The \acp{DFA} for both expressions can be seen in Figure~\ref{fig:inclusion_automaton}. 
The state numbering for the automaton in Figure~\ref{fig:inc_a2}  has been shifted for clear identification.

We begin by pushing the combined state $(1, 6)$, i.\,e., beginning state of the automaton in Figure~\ref{fig:inc_a1} and beginning state of the automaton in Figure~\ref{fig:inc_a2}. 
When popped, we can traverse from $(1, 6)$ to $(3, 9)$ with the transition $b$ and to $(2, 7)$ with the transition $a$. 
Neither of these states have two accepting states so we mark $(1, 6)$ as completed and continue. 
Next, we pop $(2, 7)$. 
Reachable from here are $(4,9)$ with $a$ and $(5,8)$ with $b$. 
Again, we have not reached a state with two accepting states, so we mark $(2,7)$ completed and continue. 
From $(5, 8)$, we can reach $(4,9)$ and $(5,9)$, as no accepting state was found we mark $(5, 8)$ completed and continue. 
Next, $(5, 9)$ is only followed by $(4,9)$ and $(5,9)$. As no accepting state was found we mark $(5, 9)$ completed and continue. 
As we can only reach $(4,9)$ and $(5,9)$ from $(4, 9)$ again, we mark $(4, 9)$ as completed. 
The last unmarked state in our stack is $(3, 9)$. 
However, we can not reach any unmarked states from here.
Now we have traversed the whole combined automaton without explicitly creating it and have not found a two-fold accepting state. 
Therefore, \regex{ab} $\subseteq$ \regex{[a-b](a|b)*} holds (see Algorithm \ref{alg:optimized_inclusion} in Section \ref{sec:algdesc}).

\subsection{Summary}
We discussed what $R^-$ rules are and how they benefit our active wrapper learning, improving performance, and usability.
Reducing the size of the $R^-$ rule set can reveal common patterns and provide guidelines for future $R^-$ rule creation.

Second, we present the procedure for regular expression inclusion.
Our algorithm, adapted from \citet{ChenX20}, takes two Python regular expressions and formalizes them.
We show how standard Python regular expressions can be converted to equivalent expressions following the formal definition.
An NFA is generated for each regular expression before being converted into a DFA.
A combined automaton is built using the complement of the presumed larger DFA and leaving the other as is.
The two automata are then tested for inclusion by iterating through the combined automatons' states, following the algorithm given by \citet{ChenX20}.

\section{Experimental Apparatus}
\label{sec:experiments}
In this section, we discuss the datasets used in this paper, describe the preprocessing steps, and the experimental procedure. 
Followed by a discussion of the measures used during the experiments.
\subsection{Datasets}
The datasets used in this paper are divided into two categories: rule sets and scientific papers.

\subsubsection{Rule set}\label{sec:data_ruleset}
We use the rule set created in the original STEREO paper using \ac{CORD}.
This dataset contains the rule set on which minimal rule set analysis is performed. 
Additionally, this rule set is used for domain comparison.
The dataset consists of in total $1,510$ rules, divided into $85\ R^+$ and $1,425\ R^-$ rules.
The rule set of STEREO is manually created using an active wrapper induction learning.
Each rule as an incremental ID (determined at the time of creation) and its corresponding regular expression.
In the inductive wrapper learning, a new rule is created if the regarded sentence is not covered by any existing rules.
This new rule is then appended to the existing set of rules, and the next higher ID is assigned to this rule (IDs are incremented by one).
Implicitly, a higher rule ID means that the rule has been added later in the process of applying the active wrapper.

It is unlikely that the $85\ R^+$ rules can be optimized greatly, as these rules match a variety of statistics. 
Statistic types are distinguished by letter, e.\,g., $U$ for Mann-Whitney-U tests vs. $t$ for $t$-test.
Thus, $R^+$ rules can most likely only include rules of the same statistic type. 
However, rules for the same statistic type primarily capture different cases, \eg different statistic values missing.
On the other hand, $1,425\ R^-$ rules can be optimized to possibly largely improve time performance and maintainability, as well as showing common patterns used to identify sentences as non-statistic.

\subsubsection{Scientific Papers}
\label{sec:scientific-papers}
These datasets provide the raw data we use for statistics extraction.
An overview of documents and sentences of these datasets can be seen in Table \ref{tab:dataset_stats}.
\paragraph{\acl{CORD}~\cite{Wang2020}}
This is the original dataset used in STEREO that can be used to evaluate the minimal rule set.
This dataset contains more than $100,000$ papers on COVID-19, SARS-CoV-2, and all coronaviruses in general.
In STEREO, the date of access is given as 21st September 2020. The CORD-19 dataset version 52\footnote{\url{https://www.kaggle.com/datasets/allen-institute-for-ai/CORD-19-research-challenge/versions/52}} (date of publication: 2020-09-21, date of access: 2022-07-11) is a very close match, which we use for comparison. 
Note, that our version is not an exact copy and contains a few more papers than the original, which leads to some slight deviations in the extraction results, i.\,e., a direct comparison is not possible.

\paragraph{arXiv Dataset~\cite{clement2019}} 
A collection\footnote{\url{https://www.kaggle.com/datasets/Cornell-University/arxiv}} (date of access: 2022-07-11) of more than 1.7 million STEM papers. 
The dataset comes with a JSON file containing metadata (author, title, category, etc.) that can be used to filter papers. 
We focus on \ac{HCI}. 
\ac{HCI} studies the use of technology, focusing on the interface between people, and computers~\cite{DBLP:books/el/LFH2017}. 
Thus, \ac{HCI} is a very promising non-medicine domain for publishing studies and the corresponding statistics. There are $9,730$ papers tagged with the ``cs.HC'' (\ac{HCI} tag on arxiv.org) category, but we focus only on papers with the HCI tag as the primary tag.
We also only use papers that were provided as both LaTeX and PDF files.
This is needed for a fair comparison of the statistics extraction on both file types.
With these restrictions, $4,023$ papers remain. 
For more information, see Appendix~\ref{sup:paper_information}.

\begin{table}[t]
    \centering
    \caption[Overview of datasets used]{An overview of the number of documents and sentences in the datasets used. Only sentences containing at least one digit were counted as only those potentially contain a statistics.}
    \label{tab:dataset_stats}
    \begin{tabular}{|l|r|r|} \hline
        \multicolumn{1}{|c|}{Dataset}   & \multicolumn{1}{c|}{\# of Documents} & \multicolumn{1}{c|}{\# of sentences}  \\\hline
        \ac{CORD}                       & $110,427$                             & $9,393,662$     \\
        \ac{HCI}                        & $4,023$                               & $222,544$       \\\hline
    \end{tabular}
\end{table}

\subsection{Preprocessing}\label{sec:preprocessing}
For the rule set inclusion, we transform the regular expressions given in the Python format into the formal definition, as described in Section~\ref{sec:algdesc}.
The Python syntax is too versatile, so this transformation into a formal, stricter expression is required to build the automata.

For the extension to the HCI domain, the papers are given as \texttt{.tar.gz} archives containing LaTeX source code. We first read the archive and extract the containing filenames using \emph{tarfile}\footnote{\url{https://docs.python.org/3/library/tarfile.html}}. 
We filter \texttt{.tex} files and use \emph{pylatexenc}\footnote{\url{https://pypi.org/project/pylatexenc/}} to read and parse the content to plain text while removing all table, figure, lstlisting, and tikzpicture environments. As in STEREO, line breaks are removed and the plain text is split into sentences using the regular expression \regex{\textbackslash .\textbackslash s?[A-Z]}. 
Every sentence that does not contain a digit is removed (cf. Table~\ref{tab:dataset_stats}).
The same process is repeated for the corresponding PDF files.
We use \emph{pdftotext}\footnote{\url{https://pypi.org/project/pdftotext/}} to easily convert the PDF files into raw text.
As before, line breaks are removed and the text is split into sentences.

\subsection{Procedure}
In the first step, we compute the minimal set of extraction rules on the existing rule set.
We do a pairwise comparison for all the rules in the rule set using Algorithm~\ref{alg:optimized_inclusion}.
This results in an array of lists containing every rule that is included by another rule.

We repeat the manual evaluation from STEREO in the \ac{HCI} domain.
Of the $4,023$ papers, we use $200$ randomly selected papers (or about $5\%$ of all papers) to learn new regular expressions to extract statistics from HCI papers.
The rule learning is applied to both LaTeX and PDF files, sampling $200$ papers for each file type.
This step generates our new rule set. 
The resulting rules are an addition to the STEREO rule set.
As we use new input formats, i.\,e., LaTeX and PDF, this will result in new format-specific rules being added, which further improves the robustness and completeness of the extraction. 

Finally, we evaluate the precision for every statistic type. 
We follow the evaluation procedure of the original STEREO paper~\cite{epp2021}.
We extract all sentences from all papers in the respective corpus that are matched by $R^+$ rules.
We then sample $200$ sentences for every statistic type, or use all extracted sentences if there were fewer than $200$ extractions, to check if the statistic types are matched and extracted correctly.
We highlight the difference in APA versus non-APA reporting.
Furthermore, we extract $200$ sentences matched by $R^-$ rules from random documents, which we did not use for rule learning.
This is to test whether $R^-$ rules do not reject statistics correctly.
Lastly, we extract $200$ sentences that were neither matched by $R^+$ nor $R^-$ to check for unrecognized statistics or parsing errors.

\subsection{Measures}
While reducing the rule set, we evaluate each rule. 
For every $R^+$ and $R^-$ rule, we measure the runtime of inclusion and the number of included rules.
We assume that every rule has a specific reason, as all rules were added due to a sentence that did or did not report a statistic. 
Nevertheless, in some cases, a general rule was added later, which includes less generalized rules, which allows to reduce the rule set size. 
A simplified example would be one rule specifying \regex{Fig\textbackslash.\textbackslash s\textbackslash d} with a later rule specifying \regex{(Fig\textbackslash.|Table|Equation)\textbackslash s\textbackslash d}~.

\vspace{0.5cm}
Similar to STEREO, our main measure is precision.
We calculate the precision by checking the $200$ rules per statistic type for false positive matches and then using the following formula:
\begin{equation}
    \text{Precision}=\frac{\text{True Positives}}{\text{True Positives}+\text{False Positives}}
\end{equation}

Furthermore, we focus on the difference in coverage of the reduced rule set in comparison to the complete, original rule set.

Finally, to gain insights on the runtime improvement of our reduced set, we evaluate the $R^-$ rules using the reduced rule set in comparison to the larger, original rule set.
We measure the time it takes to extract $200$ sentences matched by $R^-$ rules as a baseline, as here all sentences need to be checked against the list of rules.

\section{Results}
\label{sec:results}
We structure our results as in the previous sections, beginning with the rule set inclusion, then the transfer to the HCI domain, considering both PDF and LaTeX files as an input. Finally, we re-evaluate the statistics extraction using our newly created rule set.

\subsection{Rule Set Inclusion}
We ran the inclusion algorithm on $R^+$ rules and do not see much improvement.
One rule was removed, but this was an exact duplicate, most likely added by mistake.
The sub-rules of a $R^+$ rule are only used for value extraction.
Therefore, sub-rules are also removed when removing $R^+$ rules, as they serve no purpose anymore.

After running the rule inclusion algorithm for the $R^-$ rules, $483$ unique rules out of the total $1,426$ rules were included by others.
This is a reduction of about $33.8\%$.
\extended{
A distribution of the amount of included rules per rule can be seen in Figure~\ref{fig:inclusion_by_id}.}
$1,253$ rules included no other rules, $83$ included one, and $28$ included two rules.
On the other hand, $13$ rules included more than $20$ rules, and $4$ had more than $100$ rule inclusions.
These rules can be seen in Table~\ref{tab:inclusion_count}. 
Naturally, some rules were included more than once.
Figure~\ref{fig:amount_of_inclusions_graph} shows which rules were included most often.
\regex{figure \textbackslash d\{1,2\}} was the most included rule with $14$ inclusions, and \regex{table \textbackslash d+} was the second-most included rule with $13$ inclusions.

We also show the included rules sorted by rule ID (grouped in hundreds) in Figure~\ref{fig:inclusion_hist}.
We observe that many rules that were included by others had a rule ID between $400$ and $700$.
Furthermore, approximately $47\%$ of included rules had an ID below $500$ and $88\%$ had an ID below $1,000$.
In general, lower ID rules are included in higher ID rules.

\begin{table}[t]
    \centering
    \caption[Number of rules included for each rule]{Number of $R^-$ rules included for each rule. Sorted by number of included rules and rules with less than $20$ inclusions excluded.}
    \label{tab:inclusion_count}
    \begin{tabular}{lc} \hline
        Regular Expression & \# included rules \\ \hline
        \begin{minipage}{2.1in}\small\begin{verbatim}[a-zA-Z]{3,}\s?\d+[\.\,\s\dabcdef]*\end{verbatim}\end{minipage}  & $173$ \\ 
        \begin{minipage}{2.1in}\small\begin{verbatim}[a-zA-Z]{2,}\s?\d+(\.\d)?\end{verbatim}\end{minipage}  & $173$ \\ 
        \begin{minipage}{2.1in}\small\begin{verbatim}[a-zA-Z]{3,}\s?\d+[\.\,\s\d]*\end{verbatim}\end{minipage}  & $171$ \\ 
        \begin{minipage}{2.1in}\small\begin{verbatim}[a-zA-Z]{3,};?\s?\d+\end{verbatim}\end{minipage}  & $130$ \\ 
        \begin{minipage}{2.1in}\small\begin{verbatim}[a-zA-Z"]+\s?\d{1,3}$\end{verbatim}\end{minipage}  & $83$ \\ 
        \begin{minipage}{2.1in}\small\begin{verbatim}[a-zA-Z]{3,20}\s\d+(\,\d+)*(\.\d+)?\end{verbatim}\end{minipage}  & $72$ \\ 
        \begin{minipage}{2.1in}\small\begin{verbatim}[a-zA-Z]{3,20}\d+\end{verbatim}\end{minipage}  & $71$ \\ 
        \begin{minipage}{2.1in}\small\begin{verbatim}[a-zA-Z]{3,}\s-?\d+(\.\d+)?\end{verbatim}\end{minipage}  & $62$ \\ 
        \begin{minipage}{2.1in}\small\begin{verbatim}\d+(\,\d+)*(\.\d+)?\s[a-zA-Z]{3,10}\end{verbatim}\end{minipage}  & $51$ \\ 
        \begin{minipage}{2.1in}\small\begin{verbatim}[a-zA-Z]{4,10}\s\d+(,\s\d+)*\end{verbatim}\end{minipage}  & $47$ \\ 
        \begin{minipage}{2.1in}\small\begin{verbatim}\d+(\s\d+)?\s[a-zA-Z]{3,10}\end{verbatim}\end{minipage}  & $40$ \\ 
        \begin{minipage}{2.1in}\small\begin{verbatim}[a-zA-Z]{3,20}-\d+\end{verbatim}\end{minipage}  & $21$ \\ 
        \begin{minipage}{2.1in}\small\begin{verbatim}[a-zA-Z]{2}\d+(\.\d)?\end{verbatim}\end{minipage}  & $21$ \\ 
    \end{tabular}
\end{table}

\extended{
\begin{figure}
    \centering
    \includegraphics[width=\linewidth]{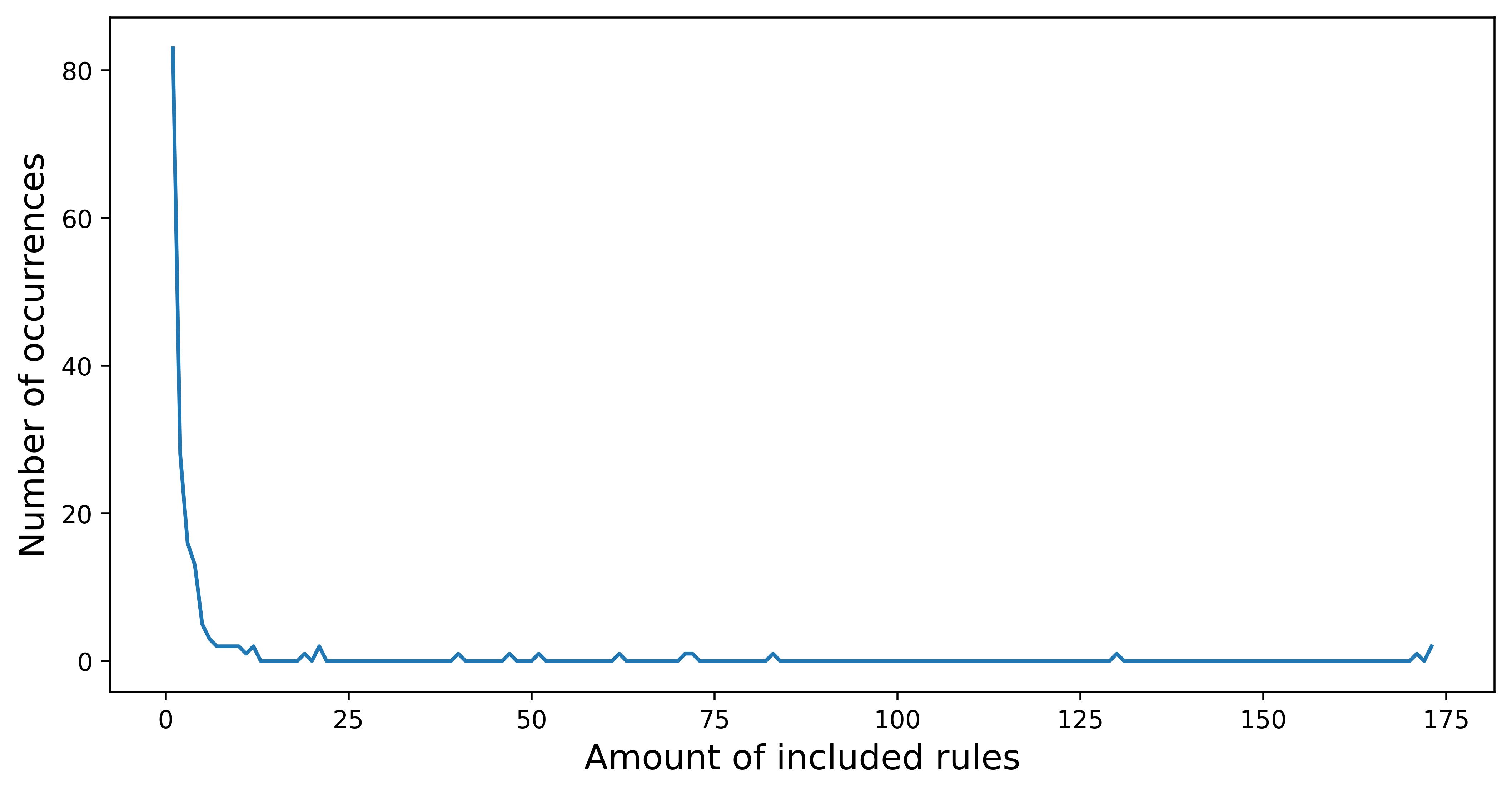}
    \caption[Amount of rules that include n other rules]{Amount of rules that include n other rules. Rules that include $0$ rules were excluded.}
    \label{fig:inclusion_by_id}
\end{figure}
}

\begin{figure}
    \centering
    \includegraphics[width=\linewidth]{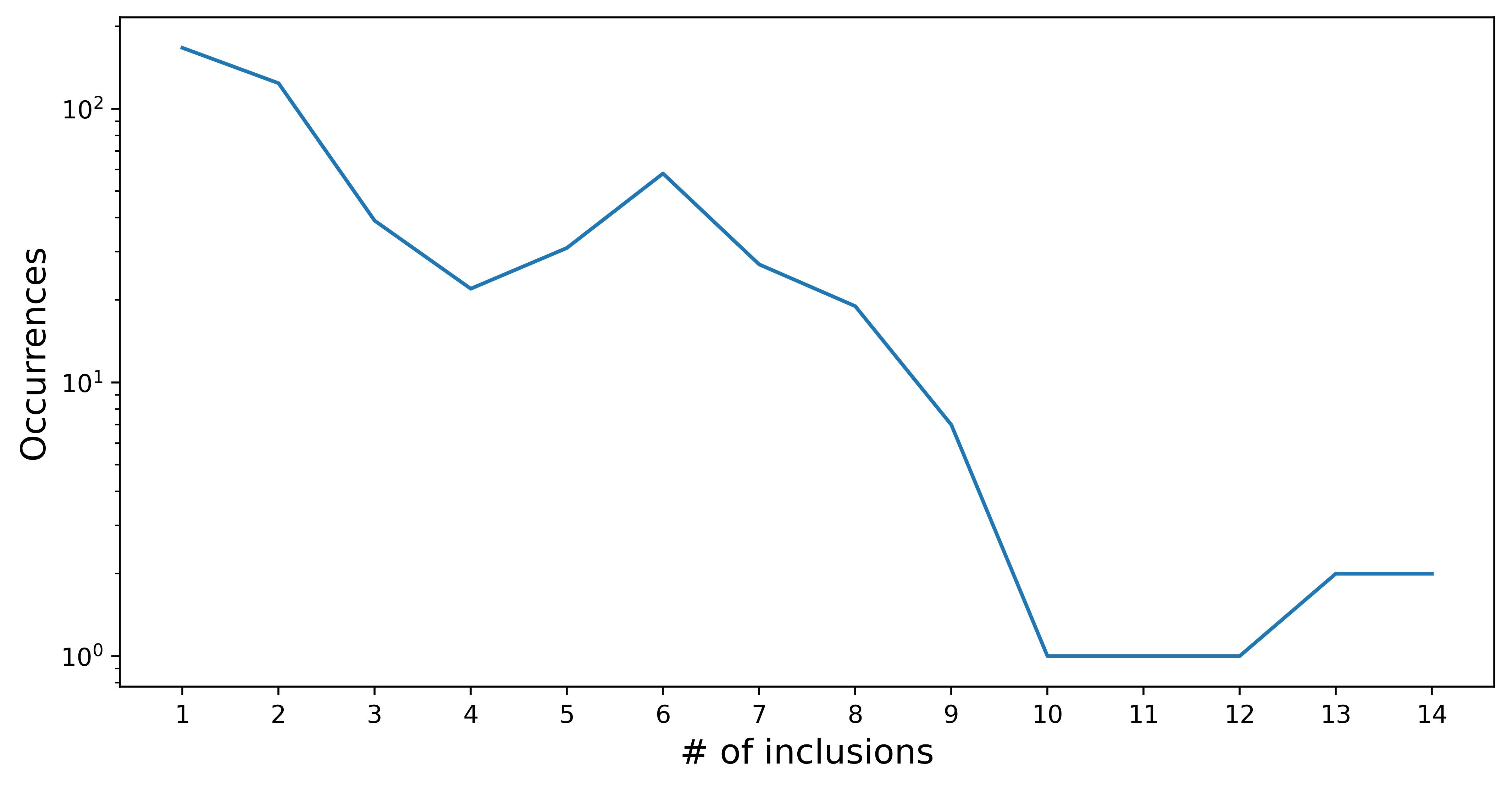}
    \caption[Number of times a rule was included by other rules]{Number of times a rule was included by other rules. The y-axis uses a logarithmic scale. $166$ rules were included once, while one rule was included $13$ and one $14$ times.}
    \label{fig:amount_of_inclusions_graph}
\end{figure}

\begin{figure}
    \centering
    \includegraphics[width=\linewidth]{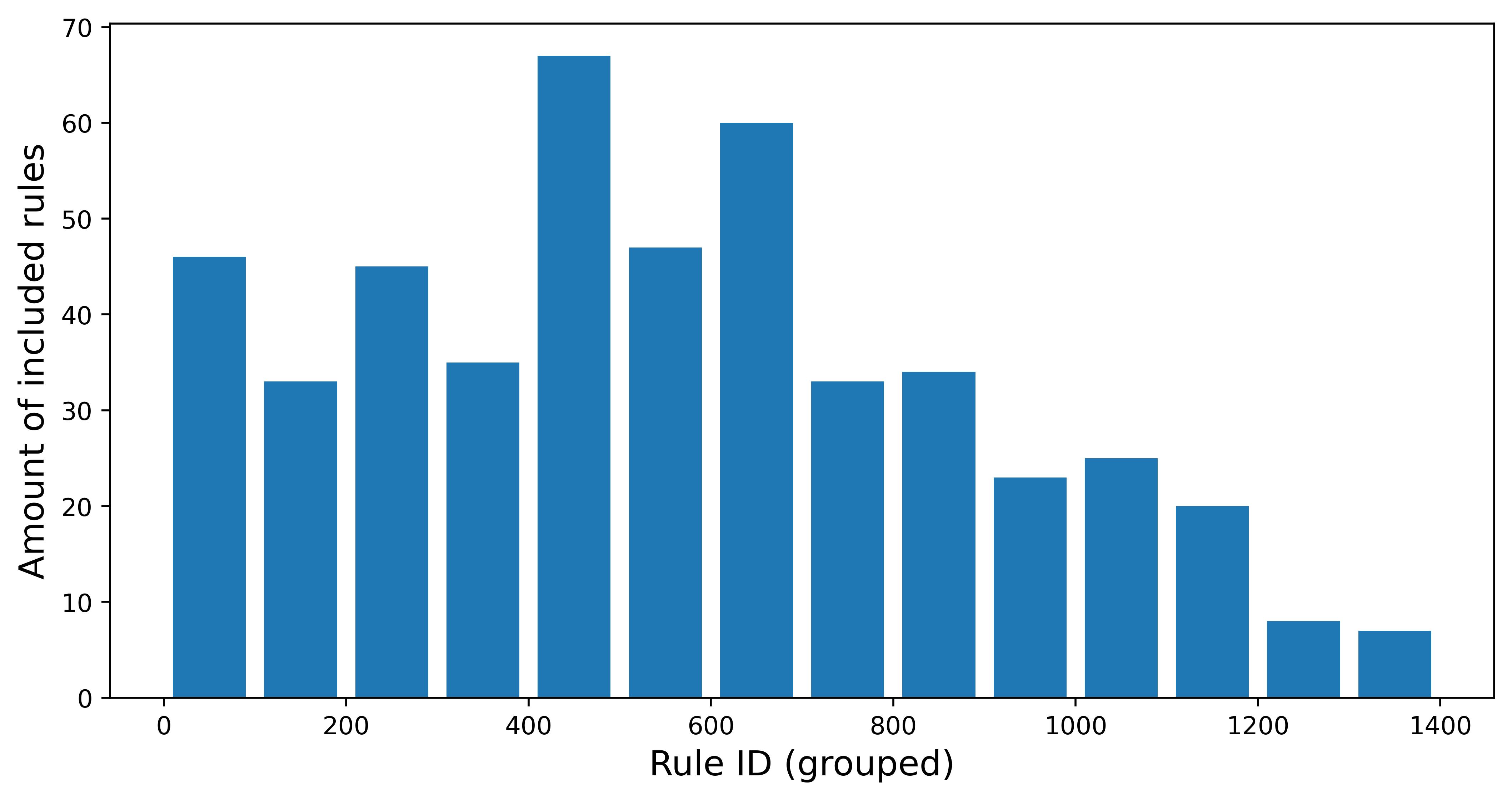}
    \caption[Amount of rules included by some other rule, ordered by rule ID (grouped by the hundreds)]{Amount of rules absorbed by rule ID, grouped by the hundreds.
    The rule ID reflects the order in which a rule was added to the rule set using STEREO's~\cite{epp2021} wrapper induction (see Section \ref{sec:data_ruleset}). When a new rule is added the ID is incremented by +1.
    \extended{See Appendix \ref{sup:deeper_analysis_inclusion_comp} for a detailed view, with no grouping.}
    }
    \label{fig:inclusion_hist}
\end{figure}

\subsection{Transfer to HCI Domain}
\paragraph{LaTeX} Using the LaTeX files, we added $13$ new $R^+$ rules and $77$ $R^-$ rules. Furthermore, $6$ previously added $R^-$ rules were changed alongside the active wrapper to be more general (see Appendix~\ref{sup:changed_rules}).
The $R^+$ rules added two new statistics types, which were not used in STEREO.
First, the Z-Test, which tests the mean of a distribution, and second, ANOVA without an $r$-value.
In the original implementation of STEREO, all ANOVA tests that did not contain a $r$-value were seen as non-\ac{APA}.
However, upon further research, we found that \ac{APA} guidelines do allow ANOVA to be reported without an $r$-value.
Therefore, when referencing the percentage of APA-conform statistics in a corpus, we mention both including and excluding ANOVA tests without an $r$-value.
For both the Z-Test and ANOVA without an $r$-value, only APA conform extraction rules were added.

\paragraph{PDF files}
For the PDF files, nine $R^-$ rules and no $R^+$ rules were added. 
We did need to add rules for page numbers and citations. 
In LaTeX files and in the CORD-19 dataset, page numbers, as well as citations, were automatically removed or never generated.
However, converted PDF files contained citations, which in turn included pages of an article in a journal or ACM identifiers.
These had a high variance of representation, making finding rules to capture them difficult.
Some examples of these variations can be seen in the following:
\begin{itemize}
    \item \texttt{Human Factors in Computing Systems. dl.acm.org, 2853–2859.}
    \item \texttt{ACM, New York, NY, USA, 285–296.}
    \item \texttt{Computer Graphics 19, 12 (2013), 2713–2722.}
    \item \texttt{Virtual Environ. 7(3), 225– 240 (1998).}
    \item \texttt{Thousand Oaks, CA, 508–510 (2007) [49]}
\end{itemize}
In the end, we added the rule \regex{\textbackslash ),\textbackslash s\textbackslash d\{1,4\}[-–]\textbackslash d\{2,4\}[.)]} to capture most, if not all, cases.
Furthermore, tables could not be removed from the \LaTeX~input, leading to some extra rules.
In general, most numbers were matched by the previously added $R^-$ and $R^+$ rules.

All in all, learning rules in the \ac{HCI} domain added $99$ rules.

\subsection{Evaluation of New Rule Set}
\paragraph{HCI Results}
Our $R^+$ rules extract the original statistic types used in STEREO~\cite{epp2021} (Pearson’s Correlation, Spearman Correlation, Student’s $t$-test, ANOVA, Mann-Whitney-U Test, Wilcoxon Signed-Rank Test, and Chi-Square Test) as well as Z-Test and ANOVA without an $r$-value, added by the previous step.

In the $4,023$ \ac{HCI} papers, the $R^+$ rules matched $6,321$ and $7,669$ sentences from the PDF and LaTeX files, respectively.
These are about $3\%$ of all sentences in the dataset for both file types.
Table~\ref{tab:stat_extraction_pdf_latex} shows all reported statistics categorized by type and whether the statistics matched \ac{APA} style or not.
For every statistic type, more statistics were extracted from LaTeX files than from PDF files.
The only exceptions are Spearman Correlation (APA) and ANOVA (non-APA), but these exceptions can be attributed to formatting errors (Spearman Correlation was marked as non-APA using LaTeX and ANOVA was marked as ANOVA without $r$-value APA).
'Other statistics' is a unspecified statistic type that includes a wide range of statistics, e.\,g., interquartile range (IQR) or Kolmogorov-Smirnov tests.
For this reason, there are no APA versions for these rules, which result in an ``N/A'' in Table \ref{tab:stat_extraction_pdf_latex}.
Using PDF files, about $26\%$ of the extracted statistics were APA conform ($9\%$ when treating ANOVA without $r$-value as non-APA).
With LaTeX files, $27\%$ of extracted statistics were APA conform ($13\%$ without ANOVA without $r$-value).

\begin{table}[t]
\caption[Extracted statistics for APA and non-APA conform reporting on \acs{HCI} papers]{Number of extracted statistics for APA and non-APA conform reporting on \ac{HCI} papers. 
Separately considering the extraction from PDF and LaTeX files.}
\label{tab:stat_extraction_pdf_latex}
\small
\begin{tabular}{l|rr|rr|}
\cline{2-5}
                                                 & \multicolumn{2}{c|}{APA conform}                        & \multicolumn{2}{c|}{non-APA conform}                    \\ \hline
\multicolumn{1}{|l|}{Statistic Type}             & \multicolumn{1}{c|}{PDF}   & \multicolumn{1}{c|}{LaTeX} & \multicolumn{1}{c|}{PDF}   & \multicolumn{1}{c|}{LaTeX} \\ \hline
\multicolumn{1}{|l|}{Student's $t$-test}         & \multicolumn{1}{r|}{440}   & $634$                        & \multicolumn{1}{r|}{38}    & $69$                         \\
\multicolumn{1}{|l|}{Pearson Correlation}        & \multicolumn{1}{r|}{48}    & $65$                         & \multicolumn{1}{r|}{76}    & 94                         \\
\multicolumn{1}{|l|}{Spearman Correlation}       & \multicolumn{1}{r|}{2}     & 1                          & \multicolumn{1}{r|}{59}    & 64                         \\
\multicolumn{1}{|l|}{ANOVA}                      & \multicolumn{1}{r|}{0}     & 0                          & \multicolumn{1}{r|}{2}     & 0                          \\
\multicolumn{1}{|l|}{ANOVA without $r$-value}    & \multicolumn{1}{r|}{$1,059$} & $1,097$                      & \multicolumn{1}{r|}{0}     & 0                          \\
\multicolumn{1}{|l|}{Mann-Whitney-U}             & \multicolumn{1}{r|}{0}     & 0                          & \multicolumn{1}{r|}{270}   & 425                        \\
\multicolumn{1}{|l|}{Wilcoxon Signed-Rank}       & \multicolumn{1}{r|}{0}     & 0                          & \multicolumn{1}{r|}{0}     & 0                          \\
\multicolumn{1}{|l|}{Chi-Square}                 & \multicolumn{1}{r|}{53}    & 85                         & \multicolumn{1}{r|}{14}    & 718                        \\
\multicolumn{1}{|l|}{Z-Test}                     & \multicolumn{1}{r|}{66}    & 195                        & \multicolumn{1}{r|}{0}     & 0                          \\
\multicolumn{1}{|l|}{Other statistics}           & \multicolumn{1}{r|}{N/A}   & N/A                        & \multicolumn{1}{r|}{$4,194$} & $4,184$                      \\ \hline
\multicolumn{1}{|l|}{Total number of statistics} & \multicolumn{1}{r|}{$1,668$} & $2,077$                      & \multicolumn{1}{r|}{$4,653$} & $5,554$                      \\ \hline
\end{tabular}
\end{table}

Using the extracted HCI sentences, we evaluate the extraction manually using a sample of $200$ sentences for every statistic type.
If fewer than $200$ sentences of a statistic were extracted, we use all extracted sentences.
For statistic types without any samples, we cannot calculate precision.

On the PDF files, the extraction precision for APA statistics was $100\%$ and ranged from $90\%$ to $100\%$ for non-APA statistics (see Table~\ref{tab:precision_pdf_latex}).
Besides that, 'Other statistics' only had a precision of $54.5\%$. 
Similarly, using the LaTeX files, we achieved $100\%$ precision for APA conform statistics and precision ranging from $89\%$ to $100\%$ otherwise.
Additionally, 'Other statistics' also only achieved a precision of $60.5\%$.

The large deviation of the precision (Table \ref{tab:precision_pdf_latex}) for 'Other statistics' compared to other statistic types can be traced back to a single $R^+$ rule.
The rule in question is \regex{\textbackslash ([P|p] \textbackslash s? <?=? \textbackslash s? \textbackslash d (\textbackslash .\textbackslash d+)?\textbackslash )}. 
This rule also captures the string \regex{(P1)}, which is not a reported statistic.
When changing the rule to \regex{\textbackslash([P|p] \textbackslash s? [<=]+ \textbackslash s? \textbackslash d (\textbackslash .\textbackslash d+)?\textbackslash)} and re-running the evaluation, only $2,337$ 'Other statistics' were extracted.
However, the precision goes up to $97.5\%$ for the PDF files.
For the LaTeX files, 'Other statistics' extractions were reduced to $2,254$, with a precision increase to $98.5\%$.

For both PDF and LaTeX files, the falsely classified Mann-Whitney-U tests were mostly Wilcoxon Signed-Rank tests and the falsely classified Pearson Correlations were mostly Spearman Correlations and vice versa.

\begin{table}[t]
\caption[Precision of extracted statistics using \acs{HCI} papers]{Precision of extracted statistics using \ac{HCI} papers. We calculate the precision on 200 samples for every statistic type (or all extracted samples, if there are less). Statistic types with no extracted samples could not be calculated.}
\label{tab:precision_pdf_latex}
\small
\begin{tabular}{l|rr|rr|}
\cline{2-5}
                                              & \multicolumn{2}{c|}{APA conform}                          & \multicolumn{2}{c|}{non-APA conform}                       \\ \hline
\multicolumn{1}{|l|}{Statistic Type}          & \multicolumn{1}{c|}{PDF}     & \multicolumn{1}{c|}{LaTeX} & \multicolumn{1}{c|}{PDF}      & \multicolumn{1}{c|}{LaTeX} \\ \hline
\multicolumn{1}{|l|}{Student's $t$-test}      & \multicolumn{1}{r|}{$100\%$} & $100\%$                    & \multicolumn{1}{r|}{$97.4\%$} & $100\%$                    \\
\multicolumn{1}{|l|}{Pearson Correlation}     & \multicolumn{1}{r|}{$100\%$} & $100\%$                    & \multicolumn{1}{r|}{$96\%$}   & $96.8\%$                   \\
\multicolumn{1}{|l|}{Spearman Correlation}    & \multicolumn{1}{r|}{$100\%$} & $100\%$                    & \multicolumn{1}{r|}{$90.7\%$} & $89\%$                     \\
\multicolumn{1}{|l|}{ANOVA}                   & \multicolumn{1}{r|}{N/A}     & N/A                        & \multicolumn{1}{r|}{$100\%$}  & N/A                        \\
\multicolumn{1}{|l|}{ANOVA without $r$-value} & \multicolumn{1}{r|}{$100\%$} & $100\%$                    & \multicolumn{1}{r|}{N/A}      & N/A                        \\
\multicolumn{1}{|l|}{Mann-Whitney-U}          & \multicolumn{1}{r|}{N/A}     & N/A                        & \multicolumn{1}{r|}{$92\%$}   & $94\%$                     \\
\multicolumn{1}{|l|}{Wilcoxon Signed-Rank}    & \multicolumn{1}{r|}{N/A}     & N/A                        & \multicolumn{1}{r|}{N/A}      & N/A                        \\
\multicolumn{1}{|l|}{Chi-Square}              & \multicolumn{1}{r|}{$100\%$} & $100\%$                    & \multicolumn{1}{r|}{$100\%$}  & $100\%$                    \\
\multicolumn{1}{|l|}{Z-Test}                  & \multicolumn{1}{r|}{$100\%$} & $100\%$                    & \multicolumn{1}{r|}{N/A}      & N/A                        \\
\multicolumn{1}{|l|}{Other statistics}        & \multicolumn{1}{r|}{N/A}     & N/A                        & \multicolumn{1}{r|}{$54.5\%$} & $60.5\%$                   \\ \hline
\end{tabular}
\end{table}

\paragraph{\ac{CORD} Results}
We ran the same evaluation on \ac{CORD}.
The results are presented in Tables~\ref{tab:stat_extraction_cord} and \ref{tab:precision_cord}.
As expected, we achieve similar results as the original STEREO paper.
The extracted sentences grouped by statistic type can be seen in Table~\ref{tab:stat_extraction_cord}.
Besides 'Other Statistics', non-\ac{APA} conform Pearson Correlations were extracted the most, with a large margin.
Of the extracted statistics, $1.8\%$ were APA conform ($0.8\%$ when treating ANOVA without $r$-value as non-APA).

We evaluate the precision, again using $200$ randomly selected sentences (see Table~\ref{tab:precision_cord}).
As for the HCI dataset, the APA-conform extractions achieved a precision of $100\%$.
For non-APA conform statistics, the precision ranged from $94.5\%$ to $100\%$.

\begin{table}[t]
\caption[Extracted statistics for APA and non-APA conform reporting on \acs{CORD} papers]{Number of extracted statistics for APA and non-APA conform reporting on \ac{CORD} papers.\extended{~Note: The number of reported statistics deviate slightly from the original STEREO paper, as a slightly different version of \ac{CORD} had to be used (see Section~\ref{sec:scientific-papers}).}}
\label{tab:stat_extraction_cord}
\small
\begin{tabular}{|l|r|r|}
\hline
Statistic Type             & \multicolumn{1}{c|}{APA conform} & \multicolumn{1}{c|}{non-APA conform} \\ \hline
Student's $t$-test         & 662                              & 210                                  \\
Pearson Correlation        & 113                              & $5,034$                                \\
Spearman Correlation       & 1                                & 551                                  \\
ANOVA                      & 0                                & 2                                    \\
ANOVA without $r$-value    & $1,239$                            & 0                                    \\
Mann-Whitney-U             & 2                                & 419                                  \\
Wilcoxon Signed-Rank       & 0                                & 0                                    \\
Chi-Square                 & 69                               & 58                                   \\
Z-Test                     & 103                              & 0                                    \\
Other statistics           & N/A                              & $114,242$                              \\ \hline
Total number of statistics & $2,189$                            & $120,516$                              \\ \hline
\end{tabular}
\end{table}

\begin{table}[t]
\caption[Precision of extracted statistics using \acs{CORD} papers]{Precision for extracted statistics using \ac{CORD} papers. We calculate the precision on 200 samples for every statistic type (or all extracted samples, if there are less). Statistic types with no extracted samples could not be calculated.}
\label{tab:precision_cord}
\small
\begin{tabular}{|l|r|r|}\hline
Statistic Type             & APA conform    & non-APA conform \\ \hline
Student's $t$-test         & $100\%$        & $97\%$     \\
Pearson Correlation        & $100\%$        & $98.5\%$   \\
Spearman Correlation       & $100\%$        & $100\%$     \\
ANOVA                      & N/A            & $100\%$      \\
ANOVA without $r$-value    & $100\%$        & N/A      \\
Mann-Whitney-U             & $100\%$        & $94.5\%$    \\
Wilcoxon Signed-Rank       & N/A            & N/A      \\
Chi-Square                 & $100\%$        & $100\%$    \\
Z-Test                     & $100\%$        & N/A      \\
Other statistics           & N/A            & $98.5\%$ \\\hline
\end{tabular}
\end{table}

\paragraph{$R^-$ rule evaluation}
We evaluate the $R^-$ rules on both HCI as well as \ac{CORD}, to test if any statistics were falsely matched by $R^-$ rules.
We sample 200 sentences matched by an $R^-$ rule from the original STEREO rule set in randomly selected papers from the respective paper corpus. 
In all scenarios, all $200$ extracted sentences were correctly identified as non-statistics.
Rerunning the experiment with the reduced $R^-$ rule set, the $200$ newly extracted sentences were also correctly identified as non-statistics.
During this step, we also measure the runtime using the \ac{HCI} dataset and compare the minimized regular expression set with the complete rule set.
As every sentence in a paper is checked against all $R^-$ rules (until a match is found), this evaluation provides a good baseline for future tasks involving the rule set.
The full rule set takes $122.4$ seconds (averaged over 5 runs), while the reduced rule set takes $84.5$ seconds (averaged over 5 runs).
This is a decrease of about $40$ seconds and is not that relevant in the case of $200$ sampled sentences.
However, when extending this to a much larger test or using the reduced set on a larger corpus, this $30\%$ difference is not negligible.

Lastly, we extract $200$ sentences that contain a number but were not matched by any $R^-$ or $R^+$ rules.
We assess whether these uncaptured sentences report a statistic or not, or whether they contain a parsing error regardless of the content (see Table~\ref{tab:uncaptured_sentences}).
$92\%$ of uncaptured sentences did not contain any uncaught statistics.
The most missed statistics ($8.5\%$) were in \ac{CORD}, while using the PDF files in the HCI domain missed the fewest ($2.5\%$).
Using the \ac{CORD} dataset also resulted in the most parsing errors. Using the LaTeX files did not result in any parsing errors.

\begin{table}
    \centering
    \caption[Evaluation of sentences not covered by $R^-$ or $R^+$ rules]{Evaluation of sentences not covered by $R^-$ or $R^+$ rules. Evaluated on a sample of $200$ sentences taken from the respective datasets.}
    \label{tab:uncaptured_sentences}
    \small
    \begin{tabular}{|l|r|r|r|}\hline
        Dataset / File Type & Statistic & Non-Statistic & Parse Error   \\ \hline
        CORD-19 / JSON      & 17        & 174           & 9             \\
        HCI     / LaTeX     & 14        & 186           & 0             \\
        HCI     / PDF       & 5         & 192           & 3             \\ \hline
    \end{tabular}
\end{table}

\section{Discussion}
\label{sec:discussion}
\subsection{Main Results}
\paragraph{Inspecting reduced $R^-$ rules}
In our experiments, we apply the rule set inclusion algorithm to reduce our rule set and thus also improve the runtime performance on tasks using the new rule set by about a third.
In Figure \ref{fig:inclusion_hist}, we see that the later a rule was added, the likelier it was to include one or more other rules.
As before, ``later'' references the point in time a rule was added during the active wrapper learning process and that a higher rule ID was assigned to it.
Furthermore, in a deeper analysis (see Appendix \ref{sup:deeper_analysis}), we found that almost all inclusions were rules included by a later rule, suggesting no unnecessary rules were added when a match was already present.
Some exceptions exist where a lower ID rule includes a rule with a higher ID. 
These few exceptions can be attributed to human error (e.\,g., adding a rule to the rule set explicitly and not due to a found sentence), as the rules in question are mostly duplicates.
This finding makes sense, as later rules were added with more background knowledge and thus were more generalized.
An example of a rule superseded by a rule created with more background knowledge is the most included rule \regex{figure \textbackslash d\{1,2\}}. 
Later rules like \regex{[fF]igure?\textbackslash s\textbackslash d+(\textbackslash s?\.\textbackslash s?\textbackslash d+)*} and \regex{(...| fig | figure | Table |...)\textbackslash s*\textbackslash d+(\textbackslash s*[\textbackslash .\textbackslash ,]\textbackslash s*\textbackslash d+)*} (shortened) include the first rule and do not only match a figure, but also tables and equations.

The structure of rules, which include many other rules, is also mostly similar.
We see that every rule, which included more than 100 other rules leveraged numbers being preceded or followed by a word, e.\,g., \regex{[a-zA-Z]\{3,\}}.
Other common patterns used often in rules were \regex{\textbackslash d(\textbackslash .\textbackslash d+)?} for a number with an optional decimal, \regex{\textbackslash s?=\textbackslash s?} for optional spaces around a symbol, \regex{[mM]} for the choice of the same letter capitalized or not, e.\,g., \regex{m} for meters, and physical units being preceded by all possible SI~\cite{Newell2019} prefixes in a character class, e.\,g., \regex{[µkmndcpfazyhMGTPEZY]?m} to capture meters with any (or none) prefixes.

\paragraph{LaTeX vs. PDF}
In total, using LaTeX files yielded more statistics than using PDF files.
In specific cases (Chi-Square Test and Spearman Correlation), using LaTeX files even extracted more samples than the CORD-19 dataset, even though CORD-19 is a much larger dataset.
Regardless of the file format used for the input, the precision is generally satisfactory.
Since all APA-conform statistics follow a very strict and well-defined pattern, it makes sense that they have a precision of $100\%$.
However, non-APA Mann-Whitney-U test rules need refinement, as in all scenarios, some Wilcoxon Signed-Rank tests were falsely identified as Mann-Whitney-U tests.

Using the \ac{HCI} dataset, about $26\%$ of the extracted statistics were APA conform. 
This is a large improvement to the $1.8\%$ of APA conform statistics in the \ac{CORD} dataset, however, \ac{CORD} is a much larger dataset.
Nonetheless, this means that the remaining $74\%$ for HCI and $98.2\%$ for \ac{CORD} of reported statistics are non-APA conform.

While using LaTeX as an input source did miss 14 statistics, we did not encounter any parsing errors, which were the case in both JSON and PDF.
Due to this better performance and the option to easily remove environments, e.\,g., tables, figures and captions, or tikzpictures, as well as the simple parsing of math formulas and special symbols, LaTeX files are the best format to use for our statistic extraction pipeline.

\subsection{Limitations and Threats to Validity}\label{sec:limitations}
We acknowledge that when transforming the given regular expression, not all Python features for regular expressions can be transformed exactly. 
For example, lookahead assertions\footnote{\url{https://docs.python.org/3/howto/regex.html\#non-capturing-and-named-groups}} or line beginning and ending symbols ($\wedge$ and $\$$) could not be represented in our formal definition exactly and were simply removed during preprocessing.
However, all rules that ignored some parts and therefore were not exact transformations were marked as such, and any potential inclusion pertaining to such a rule was checked manually. 
Not many inclusions involved such rules, however, of these inclusions, many had to be removed during the manual check, as the matches did not hold with the original Python regular expression.

Other than that, we extend the statistics extraction to the HCI domain, but our HCI paper corpus is relatively small ($4,023$ papers used) in comparison to \ac{CORD} ($>110,000$ papers).
Nonetheless, we were able to add some new statistic types and capture rules using a random sample of $200$ of papers.
Following this, one could criticize the active wrapper process itself, as only a very small portion of the corpus is used to learn rules.
However, both in the original study as well as in our experiments, we achieve high precision on the unseen test sample on both formats.

\subsection{Impact and Future Work}
In future studies, the $R^+$ rules could be refined more.
During our evaluation, we found many Wilcoxon Signed-Rank tests were falsely captured as Mann-Whitney-U tests.
While Wilcoxon Signed-Rank and Mann-Whitney-U do have similar reporting styles, a clear distinction should be made, \eg by using the surrounding words.

Furthermore, some statistic types captured by 'Other statistics' (e.\,g., Kolmogorov-Smirnov tests or odds-ratio) could be separated and adapted to their own statistic type. 
This would more clearly show the distribution of reported statistics and potentially further improve precision and robustness.

Of course, the same procedure we used can be applied to more domains to find new statistic types, improve existing statistic types, and extend the $R^-$ rules.
In Appendix \ref{sup:guidelines}, we offer some recommendations for future rule creation.
Expanding the statistic types is also motivated by the $2.5\%$ (HCI) and $8.5\%$ (CORD-19) of missed statistics in a sample of $200$ sentences (see Table \ref{tab:uncaptured_sentences}).
Now that the implementation has been modified, it can handle papers submitted in JSON, PDF, and LaTeX, which should account for the majority of input formats.

Moreover, one could adapt our regular expression inclusion algorithm to additionally check for sub-expression inclusion (i.\,e., if a regular expression is wholly included as a sub-expression of another regular expression).
Using a simple example, the expression \regex{abc} includes \regex{bc} as a sub-expression.
Although this does not directly help us to further reduce our rule set, we can use this for a similar extension.
One could develop an algorithm that, given two regular expressions, automatically generates the smallest regular expression that accepts both expressions.
With these modifications, sub-expression inclusion can be used to further reduce the rule set size.
Our example with \regex{abc} and \regex{bc} can be merged as \regex{a?bc}.

\section{Conclusion}\label{sec:conclusion}
In this paper, we practically apply an algorithm to determine the inclusion of regular expressions.
We employ the STEREO statistics extraction tool which extracts statistics from articles using a set of regular expressions.
One third of the rules are removed when our inclusion algorithm is applied to the list of 1,510 regular expressions used by STEREO.
Furthermore, we extend the rule set to the HCI domain. 
We repeat the active wrapper learning on a sample of $200$ papers, adding 13 $R^+$ and 86 $R^-$ rules.
This is only a small fraction of newly required rules compared to the $1,510$ original rules in STEREO.
We apply the extended statistics extraction rule set to the whole HCI dataset.
We find that only $26\%$ of extracted statistics were APA conform in the \ac{HCI} domain.
This is a large improvement compared to the $<2\%$ achieved on \ac{CORD}. However, still the majority of reported statistics in HCI are not reported in accordance with the APA style guide.
Additionally, we compare the use of PDF and LaTeX files as an input.
Due to better extraction precision and fewer parsing errors, we advise using LaTeX files.

For repeatability and further insights we provide our source code: \url{https://github.com/Tobi2K/BachelorThesis}
\begin{acks}
\extended{
We thank the authors of STEREO~\cite{epp2021} for providing the source code, rule set, and their expertise.
The original source code of STEREO can be found here: \url{https://github.com/Foisunt/STEREO}.
We also thank Carolin Schindler, Simon Birkholz, and Danial Podjavorsek for providing the crawler (\url{https://github.com/Data-Science-2Like/arXive-crawler}) we used to gather the arXiv papers.}
This work is co-funded under the 2LIKE project by the German Federal Ministry of Education and Research (BMBF) and the Ministry of Science, Research and the Arts Baden-Württemberg within the funding line Artificial Intelligence in Higher Education.
\end{acks}

\bibliographystyle{ACM-Reference-Format}
\bibliography{library}

\clearpage
\balance
\addtocontents{toc}{\protect\setcounter{tocdepth}{1}} 
\begin{appendices}

\section{Lessons learned}\label{sup:guidelines}
On the basis of our analysis of the rule set, we offer recommendations on how rules should be created in the future.

This section is not a complete set of guidelines one should follow, but rather a collection of common patterns and recommendations we learned during rule creation and analysis.
\subsection{General}
In general, it is important that rules are not too generic, as these will produce many false positives.
This applies to both $R^+$ and $R^-$ rules.
For example, an $R^+$ rule capturing any equals sign followed by a number is too broad.
Almost all statistics would be captured by this rule.
However, this prevents correct value extraction and a differentiation between statistic types is not applicable, defeating the purpose of extracting specific statistic types.
However, the rules should not be too specific either. 
Although this does not produce false positives, rules that only capture a single case worsen maintainability and usability, as every case would need a new rule.
We show some patterns in the following, with which one can easily create good rules.

\subsection{Leverage Contextual Information}\label{sec:context_info}
As can be seen in Table \ref{tab:inclusion_count}, most of the rules that include many others, leverage being preceded or followed by a word. 
This context can be used to rule out a sentence or number as a statistic, as statistics usually use a maximum of $2$ letters as an identifier (e.g. \regex{p} or \regex{t}). 
So capturing a number following a word, especially without an operator, rules out that the sentence contains a statistic. 

On the other hand, the context can also be used to capture (non-APA) statistics.
For example, some existing $R^+$ rules capture a sentence containing ``significance'' followed by a $p$-value.
Significance values usually only report a $p$-value, however the simple pattern \regex{p=\textbackslash d(\textbackslash .\textbackslash d+)?} also captures all $p$-values reported in any other statistic.
To prevent this, rules can be limited by only capturing the $p$-value following the word ``significance''.
An example of such a rule is \regex{significantly higher[\textbackslash sa-zA-Z]+ \textbackslash (\textbackslash s?[pP]\textbackslash s?<\textbackslash s?0?\textbackslash .\textbackslash d+\textbackslash s?\textbackslash )}.
Here, ``significantly higher'' can be followed by some other letters and then the $p$-value in parentheses afterwards.

\subsection{Use Optional Sequences}
We already mentioned some of the following patterns which use optional sequences:
\begin{itemize}
    \item \regex{\textbackslash d(\textbackslash .\textbackslash d+)?} for a number with an optional decimal
    \item \regex{\textbackslash s?=\textbackslash s?} for optional spaces around a symbol
    \item \regex{[mM]} for the choice of the same letter capitalized or not
    \item \regex{[µkmndcpfazyhMGTPEZY]?m} to capture physical units being preceded by all possible SI~\cite{Newell2019} prefixes
    \item \regex{[<>=$\leq\geq$]} (or a subset) to capture mathematical operators
    \item \regex{[a-zA-Z\textbackslash s]+} to capture a word or words. Can also be modified:
    \begin{itemize}
        \item \regex{[a-zA-Z\textbackslash s]\{,20\}} to capture up to 20 letters or white spaces
        \item \regex{[a-zA-Z\textbackslash s]\{20,\}} to capture at least 20 letters or white spaces
    \end{itemize}
\end{itemize}
These patterns can combat different formatting or spelling mistakes and should always be used, unless there is an explicit reason not to.
Using these patterns also prevents excess rule creation.

\subsection{Grouping Similar Structures}
Numbers are often presented in a similar context.
A simple example is the numbering of tables, figures, algorithms, etc.
Following Section \ref{sec:context_info}, we can use this context and further combine these into a single rule.
A shortened example is \regex{(Fig. | Figure | fig | figure | FIGURE | Table | table | ... ) -? \textbackslash s? \textbackslash d+ (\textbackslash s [.,] \textbackslash s \textbackslash d+)?}.
This rule captures figure and table references in multiple formats.
For example, \regex{Table 1}, \regex{FIGURE 2,1}, \regex{fig-3.6}, and more, are all captured.
Another non-statistic example of this being applied to are page number references.
Here we group the options \regex{(pages|page|pp\textbackslash .|pp)} in one rule.

This grouping can also be applied to capturing statistics. 
For example, STEREO uses a rule to capture Mann-Whitney-U tests that requires the test statistics to be preceded by the name ``Mann-Whitney-U''.
To be more lenient, many options are possible: \regex{(Mann-Whitney-U test|MWU test| mwu test|Mann-Whitney test)}.

\subsection{Greek Letters and PDF Formatting}
Furthermore, it is advisable to keep different formats in mind when creating rules.
An example used in STEREO is the Chi-Square test. 
In \ac{CORD}, the used documents always converted a $\chi$ to an \texttt{x} or \texttt{X}.
Therefore, rules capturing Chi-Square tests only used the capture group \regex{[xX]}.
However, LaTeX files use \verb!$\chi$! to create the symbol $\chi$ symbol.
The LaTeX conversion we use can interpret this command directly and convert the symbol as a Unicode symbol (U+03C7), instead of an \texttt{X}.
Thus, we advise to add Greek letters as an option, if applicable, especially for $R^+$ rules.

Moreover, many scientific papers are written using LaTeX and converted to PDF.
Some characters undergo specific formatting during this conversion.
A common occurrence is using a hyphen.
Depending on the formatting and font, a hyphen can be written as - or \mbox{--} (Unicode: U+002D vs. U+2013).
Other less common hyphens are \textemdash\ (U+2014) and $-$ (Math-mode minus, U+2212).
When writing a rule containing a hyphen, e.\,g., ranges of numbers, we recommend using a capture group containing at least the first two (\regex{[-\mbox{--}]}), if not all four (\regex{[-\mbox{--}\textemdash$-$]}) alternatives.

\section{Deeper Analysis}\label{sup:deeper_analysis}
This section presents some further research conducted.

\subsection{Inclusion Comparison by Rule ID}\label{sup:deeper_analysis_inclusion_comp}
\begin{figure}[ht]
    \centering
    \includegraphics[width=\linewidth]{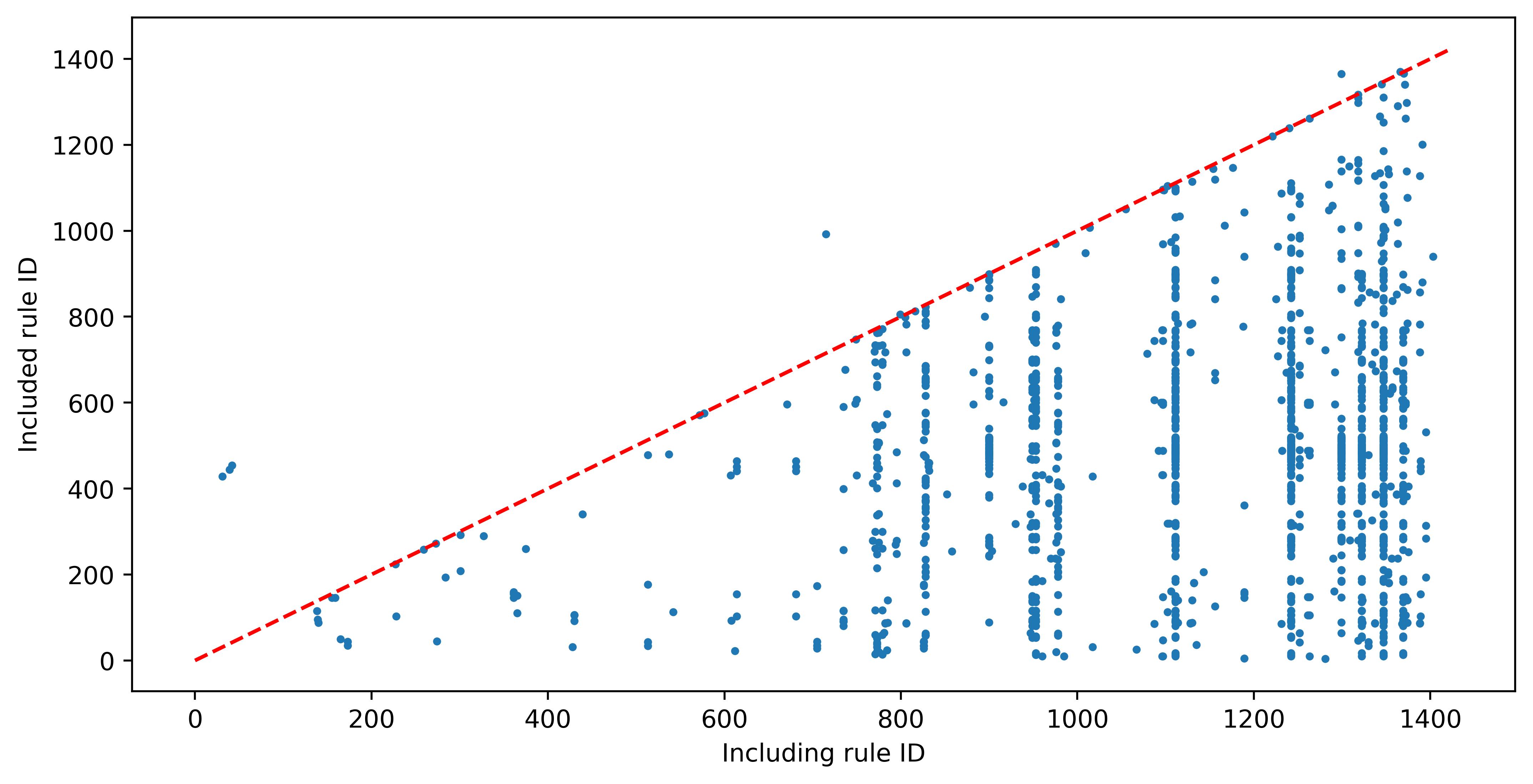}
    \caption[A comparison of rules with their respective inclusions by rule ID]{A comparison of rules with their respective inclusions by rule ID. The rule ID represents the order in which rules were added in STEREO~\cite{epp2021}. The red straight line shows the angle bisector, where every rule would include themselves. The x-axis defines the including rule ID, i.\,e., the rule that included others, while the y-axis shows the rule IDs of rules that were included.}
    \label{fig:inclusion_comparison}
\end{figure}

In Figure \ref{fig:inclusion_hist}, we show the amount of rules removed, grouped by rule ID.
Figure \ref{fig:inclusion_comparison} shows a more detailed view, where we split rules.
We observe, that almost all rule inclusion were cases where a higher rule ID was included a lower rule ID.
Some exceptions exist, however these are mostly duplicates.

\subsection{Extraction with Original Rules}

To understand how well the original rules transfer to a new domain, we reran the extraction on HCI without the rules we added during active wrapper learning.
During active wrapper learning we added rules for $t$-test (APA and non-APA), Pearson Correlation (APA and non-APA), Spearman Correlation (non-APA), Mann-Whitney-U (non-APA), and Chi-Square (APA and non-APA).
These differences can be seen in Table \ref{tab:stat_extraction_pdf_latex_original_rules} in comparison to Table \ref{tab:stat_extraction_pdf_latex}.
\begin{table}[ht]
\caption[Number of extracted statistics for APA and non-APA conform reporting on \acs{HCI} papers using only original rules]{Number of extracted statistics for APA and non-APA conform reporting on \acs{HCI} papers using only original rules. Split into extraction from PDF and LaTeX files.}
\label{tab:stat_extraction_pdf_latex_original_rules}
\small
\begin{tabular}{l|cc|cc}
\cline{2-5}
                           & \multicolumn{2}{c|}{APA conform}   & \multicolumn{2}{c}{non-APA conform} \\ \hline
Statistic Type             & \multicolumn{1}{c|}{PDF}   & LaTeX & \multicolumn{1}{c|}{PDF}  & LaTeX  \\ \hline
Student's $t$-test         & \multicolumn{1}{c|}{377}   & 634   & \multicolumn{1}{c|}{18}   & 38    \\
Pearson Correlation        & \multicolumn{1}{c|}{42}    & 65    & \multicolumn{1}{c|}{67}   & 94    \\
Spearman Correlation       & \multicolumn{1}{c|}{2}     & 1     & \multicolumn{1}{c|}{51}   & 64    \\
ANOVA                      & \multicolumn{1}{c|}{0}     & 0     & \multicolumn{1}{c|}{5}    & 0     \\
Mann-Whitney-U             & \multicolumn{1}{c|}{0}     & 0     & \multicolumn{1}{c|}{34}   & 42    \\
Wilcoxon Signed-Rank       & \multicolumn{1}{c|}{0}     & 0     & \multicolumn{1}{c|}{0}    & 0     \\
Chi-Square                 & \multicolumn{1}{c|}{1}     & 0     & \multicolumn{1}{c|}{0}    & 0     \\
Other statistics           & \multicolumn{1}{c|}{N/A}   & N/A   & \multicolumn{1}{c|}{$4,124$} & $4,141$  \\ \hline
Total number of statistics & \multicolumn{1}{c|}{422}  & 700  & \multicolumn{1}{c|}{$4,299$} & $4,379$ \\ \hline
\end{tabular}
\end{table}

\section{Thompson's Construction}\label{sup:tc_example}
\begin{figure*}
    \centering
    \includegraphics[width=\linewidth]{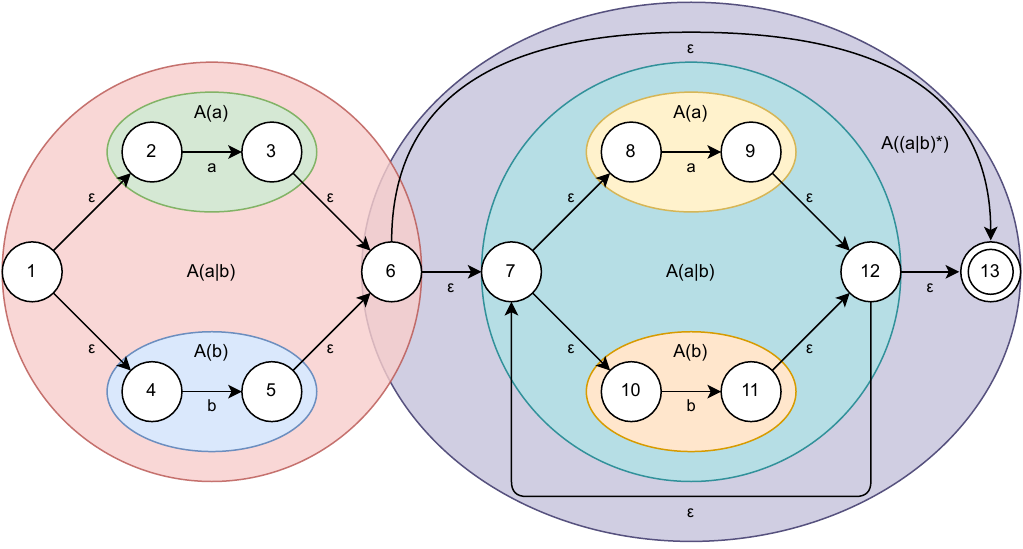}
    \caption[Larger image of Thompson's construction for \texttt{ba|ab|*\&}]{Larger image of Thompson's construction for \regex{ba|ab|*\&} (originally \regex{[a-b](a|b)*})}
    \label{fig:tc_example_large}
\end{figure*}

See Figure~\ref{fig:tc_example_large}. The colored circles represent the default automata used in Thompson's construction, cf. Figure \ref{fig:tc_basic_nfas}.

\extended{
\section{Unoptimized Automaton Inclusion}\label{sup:automaton_inclusion}

Algorithm~\ref{alg:aut_inclusion} is the original, unoptimized version of Algorithm~\ref{alg:optimized_inclusion} presented by \citet{ChenX20}.
As one can see in Algorithm~\ref{alg:optimized_inclusion}, the optimized version does not explicitly create any cross product of states.
Instead, we iterate state-by-state using the same transition letter for both automata, keeping track of our current position with a tuple $(p,q)$.
The algorithm terminates as soon as we reach a goal state in both automata, returning earlier than without the optimization.
Only if the automata are in an inclusion relation (i.\,e., no goal state is found), the graph is searched completely, resulting in the same performance as without the alteration.

This improvement benefits us as many regular expressions have a very specific use-case.
These regular expressions most likely cannot be generalized and included by some other rule, so an early exit is possible.
\begin{algorithm}[h]
\SetNoFillComment
\LinesNumbered
\caption{Automaton inclusion as proposed by \citet{ChenX20}}
\label{alg:aut_inclusion}
\KwData{Two \acp{DFA} $A_1$ and $A_2$}
\KwResult{Boolean whether $A_2 \subseteq A_1$}
$A' \gets computeComplement(L(A_1))$ \tcp*{Step 2}
$B \gets createByLanguage(L(A_2) \cap L(M'))$ \tcp*{Step 3}
\tcc{Step 4}
Construct Graph $G = (Q_B \times \delta_B$)\;
\tcc{$Q_B$ is equivalent to $Q_{A_2} \times Q_{A'}$}
\eIf{$G$ contains path from start state to accepting state}
{
    \Return $FALSE$\;
}{
    \Return $TRUE$
}
\end{algorithm}
}

\section{Information on arXiv papers}\label{sup:paper_information}
Of the original $9,730$ papers tagged with ``cs.HC'', $6,110$ papers had ``cs.HC'' as the primary tag, of which $5$ were inaccessible on the date of access (2022-07-11).
This usually means that papers have been withdrawn by the authors or were removed from arXiv.
Of these $6,105$ papers, $2,037$ were only available as PDF, $4,067$ had source code, i.\,e., LaTeX files, and one paper was only available as a \texttt{.docx} document.
We only use papers that were provided as both LaTeX and PDF files.
This is needed for a fair comparison of the statistics extraction on both file types.
$44$ papers given as source code did not have PDF files, which means $4,023$ papers remain.

\section{Changed rules}\label{sup:changed_rules}
While learning new rules for the HCI domain, we changed six previous $R^-$ rules.
These changes were made, as we had knowledge of previously added rules.
We only broadened rules, allowing more cases, for example additional white spaces.
The following changes have been made:
\begin{itemize}
    \item Allow \emph{2D} and \emph{3D} with both capital and lowercase letter \\ \verb!\s[2|3]D\s! \\ $\longrightarrow$ \\ \verb!\s[2|3][d|D]\s!
    \item Allow \emph{2D} and \emph{3D} with both capital and lowercase letter \\ \verb!\s[2|3]D\d\s! \\ $\longrightarrow$ \\ \verb!\s[2|3][d|D]\d\s!
    \item Allow any white space instead of only spaces \\ \verb!\s\d\) [a-zA-Z]+! \\ $\longrightarrow$ \\ \verb!\s\d\)\s{1,2}[a-zA-Z]+!
    \item Allow additional operators \\ \texttt{[a-zA-Z]\textbackslash s?[a-zA-Z]\textbackslash s?[a-zA-Z]\textbackslash s?[a-zA-Z] \textbackslash s? [a-zA-Z]\textbackslash s?[$\approx$=]\textbackslash s?\textbackslash d+} \\ $\longrightarrow$ \\ \texttt{[a-zA-Z]\textbackslash s?[a-zA-Z]\textbackslash s?[a-zA-Z]\textbackslash s?[a-zA-Z] \textbackslash s? [a-zA-Z]\textbackslash s?[$\approx$=>]\textbackslash s?\textbackslash d+}
    \item Allow additional white spaces \\ \verb!\d+(\.\d+)?,\s[a-zA-Z]{3,}! \\ $\longrightarrow$ \\ \verb!\d+(\.\d+)?,\s+[a-zA-Z]{3,}!
    \item Allow optional negative values in lists \\ \verb!\)\s?(,|or)?\s?\d\s?\(! \\ $\longrightarrow$ \\ \verb!\)\s?(,|or)?\s?-?\d{1,2}\s?\(!
\end{itemize}

\end{appendices}

\end{document}